\documentclass[11pt,a4paper]{article}
\pdfoutput=1
\usepackage{jheppub}

\usepackage{amsmath}
\usepackage{amssymb}
\usepackage{epsfig,multicol,bbm}
\usepackage{graphicx}
\usepackage{multirow}
\usepackage{bbold}
\usepackage[mathscr]{euscript}

\newcommand{\app}[1]{Appendix~\ref{#1}}

\newcommand{\be}[0]{\begin{equation}}
\newcommand{\ee}[0]{\end{equation}}

\newcommand{\eq}[1]{Eq.~\eqref{eq:#1}}

\renewcommand{\sec}[1]{Sec.~\ref{sec:#1}}

\newcommand{\bea}{\begin{eqnarray}}
\newcommand{\eea}{\end{eqnarray}}

\def\nslash{n\!\!\!\slash}

\newcommand{\nn}{\nonumber}

\newcommand{\as}{\alpha_s}

\newcommand{\cO}{\mathcal{O}}

\title{Analytic and Monte Carlo Studies of Jets with Heavy Mesons and Quarkonia}

\author[a]{Reggie Bain,}
\author[b]{Lin Dai,}
\author[c]{Andrew Hornig,}
\author[b]{Adam K. Leibovich,}
\author[a]{Yiannis Makris}
\author[a]{and Thomas Mehen}

\affiliation[a]{Department of Physics, Duke University,\\ Science Dr., Durham, NC 27708, USA}
\affiliation[b]{Pittsburgh Particle Physics Astrophysics and Cosmology Center (PITT PACC), Department of Physics and Astronomy, University of Pittsburgh,
\\ 3941 O'Hara St., Pittsburgh, PA 15260, USA}
\affiliation[c]{Theoretical Division T-2, Los Alamos National Laboratory,\\ Los Alamos, NM, 87545, USA}

\emailAdd{rab59@duke.edu}
\emailAdd{lid33@pitt.edu}
\emailAdd{ahornig@lanl.gov}
\emailAdd{akl2@pitt.edu}
\emailAdd{yiannis.makris@duke.edu}
\emailAdd{mehen@phy.duke.edu}

\abstract{
We study jets with identified hadrons in which a family of jet-shape variables called angularities are measured,
extending the concept of fragmenting jet functions (FJFs) to these observables.
FJFs determine the fraction of energy, $z$, carried by an identified hadron in a jet
with angularity, $\tau_a$. The FJFs are convolutions of fragmentation functions (FFs),
evolved to the jet energy scale, with perturbatively calculable matching coefficients. Renormalization group equations
are used to provide resummed calculations with next-to-leading  logarithm prime (NLL') accuracy.
We apply this formalism to
two-jet events in $e^+ e^-$  collisions with $B$ mesons in the jets, and three-jet events in which a $J/\psi$ is produced in the
gluon jet. In the case of $B$ mesons, we use a phenomenological FF
extracted from $e^+ e^-$ collisions at the $Z^0$ pole evaluated at the scale $\mu = m_b$. For events with $J/\psi$, the
FF can be evaluated in terms of Non-Relativistic QCD (NRQCD) matrix elements at the scale $\mu =2 m_c$.
The $z$ and $\tau_a$ distributions from our NLL' calculations are compared with predictions from monte carlo event generators.
While we find consistency between the predictions for $B$ mesons and the
$J/\psi$ distributions in $\tau_a$, we find the  $z$ distributions for $J/\psi$ differ significantly.
We describe an attempt to merge PYTHIA showers with NRQCD FFs that gives good
agreement with NLL' calculations of the $z$ distributions.
}

\preprint{LA-UR-16-21884}

\begin{document}
\maketitle
\noindent


\section{Introduction}
\label{sec:intro}
The study of jets and heavy flavor continues to play an important role at the Large Hadron Collider (LHC)
and many other high energy and nuclear experiments. Such studies are essential for testing our
understanding of Quantum Chromodynamics (QCD) and for calculating backgrounds in searches for new physics.
In this paper we calculate cross sections for $e^+e^-$ to jets, where one of the jets contains a hadron
with either open or hidden heavy flavor. In particular, we will derive factorization theorems and
perform analytical Next-to-Leading-Log prime (NLL') resummation\footnote{NLL' includes NLL resummation  for each function in the factorization theorem,
where all functions are computed to NLO~\cite{Almeida:2014uva}.} for these cross
sections using renormalization group (RG) techniques. We will also compare our results with monte carlo simulations of the same cross sections.

Recently, there has been considerable interest in cross sections of this type~\cite{Procura:2009vm,Liu:2010ng,Jain:2011xz,Jain:2011iu,Procura:2011aq,Jain:2012uq,Bauer:2013bza,Baumgart:2014upa,Kaufmann:2015hma,Chien:2015ctp}.
Ref.~\cite{Procura:2009vm} demonstrated that the cross section for producing a jet with an identified hadron can be
determined using a distribution function called the fragmenting jet function (FJF). FJFs are in turn related to the
more commonly studied fragmentation functions (FFs) by a matching calculation at the jet energy scale. This implies that cross sections for
jets with an identified hadron provide a new arena  to measure FFs, which are more commonly extracted from the semi-inclusive cross section
$e^+ e^- \to H + X$. Especially  important is that this provides an opportunity to extract gluon FFs \cite{Kaufmann:2015hma,Chien:2015ctp},
since quark FFs are more readily studied in $e^+ e^-  \to H + X$.
In addition, it was recently shown in Ref.~\cite{Baumgart:2014upa} that since the FFs for quarkonia production
can be calculated in the Non-Relativistic Quantum Chromodynamics (NRQCD) factorization formalism~\cite{Bodwin:1994jh}, FJFs can be used to make novel tests of quarkonium production theory.

The FJF was first introduced in Ref.~\cite{Procura:2009vm} whose main results can be summarized as follows:
\begin{itemize}
\item  A factorization theorem for a jet with an identified hadron, $H$, is obtained from the
factorization theorem for a jet cross section by the replacement
\bea\label{eq:ReplacementRule}
J_i(s,\mu) \to \frac{1}{2(2\pi)^3}{\cal G}^H_i(s,z,\mu) dz ,
\eea
where $J_i(s,\mu)$  is the jet function for a jet with invariant mass $s$ initiated by parton $i$, and the
renormalization scale is $\mu$. The FJF, denoted ${\cal G}^H_i(s,z,\mu)$, additionally depends on
the fraction $z$ of the jet energy that is carried by the identified hadron. These functions implicitly depend on the jet
clustering algorithm and cone size $R$ used to define the jets. It is also possible to
define jet functions and FJFs that depend on the total energy of the jet rather than the invariant mass~\cite{Procura:2011aq}.

\item The FJFs, ${\cal G}^H_i(s,z,\mu)$, are related to the well-known FFs, $D^H_i(z,\mu)$, by the formulae
\bea\label{eq:GintermsofD}
{\cal G}^H_i(s,z,\mu) =  \sum_{j} \int_z^1 \frac{d z'}{z'}{\cal J}_{ij}(s,z',\mu)  D^H_j(z/z',\mu)  +{\cal O}\left(\Lambda^2_{\rm QCD}/s\right) \, ,
\eea
where the coefficients ${\cal J}_{ij}(s,z,\mu)$ are perturbatively calculable matching coefficients whose
large logs are minimized at the jet scale, $s$, and are calculated to NLO in
Ref.~\cite{Jain:2011xz}. For heavy quarks the ${\cal J}_{ij}(s,z,\mu)$ have been calculated to $O(\alpha_s^2)$
in Ref.~\cite{Bauer:2013bza}.
\item These matching coefficients obey the sum rule
\bea\label{sumrule}
J_i(s,\mu) =\frac{1}{2(2\pi)^3} \sum_j \int_0^1dz z {\cal J}_{ij}(s,z,\mu) \, .
\eea
\end{itemize}
The properties of FJFs were further studied in Refs.~\cite{Liu:2010ng,Jain:2011xz,Jain:2011iu,Procura:2011aq,Jain:2012uq}. These papers focused
on the FJFs for light hadrons such as pions. FJFs for particles with a single heavy quark were
studied in Ref.~\cite{Bauer:2013bza} and FJFs for quarkonia were calculated in Ref.~\cite{Baumgart:2014upa}.

One important goal of this work is to generalize FJFs to jets in which the angularity is measured.
The angularity, denoted  $\tau_a$,  is defined as \cite{Berger:2003iw}
\bea
\label{eq:tauadefn}
\tau_a = \frac{1}{\omega}\sum_i (p_i^+)^{1-a/2} (p_i^-)^{a/2} \, ,
\eea
where the sum is over all the particles in the jet, and $\omega = \sum_i  p_i^-$ is the large light-like
momentum of the jet. The angularity should be viewed as a generalization of the  invariant mass squared
of the jet since $s=\omega^2 \tau_0$. We calculate the matching coefficients appropriate for jets in
which the angularity has been measured, denoted ${\cal J}_{ij}(\tau_a, z,\mu)$, and
verify the $s \to \tau_a$ generalization of the sum rules in Eq.~(\ref{sumrule}) in \app{match} of this paper.
The other goal of this work is to study the $z$ and $\tau_a$ dependence of the cross section
for jets with identified heavy hadrons in $e^+e^-$ collisions and compare our analytical results to monte carlo simulations.
We will do this for two-jet events in which $e^+e^- \to b\bar{b}$ is followed by fragmentation to $B$ mesons.
We will also study three-jet events with $e^+e^- \to b\bar{b}g$ followed by the gluon fragmenting to
a jet with a $J/\psi$. At the LHC we expect high energy gluons fragmenting to a jet with $J/\psi$ to be an important
production mechanism of $J/\psi$ at high $p_T$ and Ref.~\cite{Baumgart:2014upa} showed this process is sensitive to the mechanisms
underlying quarkonium production. The study in this paper will allow comparison of analytic calculations with monte carlo simulations of gluons fragmenting to $J/\psi$ in jets.
In order for this cross section to be physically observable one would either include quarks and antiquarks
fragmenting to jets with $J/\psi$ or one would have to ensure experimentally that  the $J/\psi$
came from the gluon jet in the three-jet event, which could be possible if the other jets are $b$-tagged.

In Section~\ref{sec:FJFs}, we discuss the basics of FJFs for events containing jets
where the angularity of the one of the jets is probed. We review various properties of FJFs and their relationship with
the more commonly studied FFs. We also present our results for the matching coefficients $\mathcal{J}_{ij}(\tau_a,z,\mu)$
for jets with measured angularities. Further details of that calculation can be found in \app{match}. In Section~\ref{sec:2jets}, we present
our results for the NLL' cross section for $e^+e^- \to 2$ jets where one of the jets contains a
$B$ meson and the angularity of that jet is measured. We find reasonable agreement in both $z$ and $\tau_a$ distributions
between our analytic  calculations and monte carlo simulations
performed using Madgraph~\cite{Alwall:2014hca} $+$ PYTHIA~\cite{Sjostrand:2006za, Sjostrand:2014zea} and Madgraph $+$ HERWIG~\cite{Bahr:2008pv}.
In Section~\ref{sec:3jets}, we show similar comparisons of analytic versus
monte carlo calculations for the cross section for $e^+e^- \to 3$ jets where one of the jets contains a $J/\psi$ created via gluon
fragmentation. In this case the $\tau_a$ distributions for the jet are in good agreement, but the monte carlo predictions for the $z$
distributions are inconsistent.  We believe that this is due to PYTHIA's modeling of radiation from color-octet states that produces a harder $z$
distribution than the analytic calculations. In an effort to improve the consistency
between NLL' and monte carlo calculations, we turn off hadronization in PYTHIA and then convolve the distribution of momenta of the gluons within a jet with the NRQCD color-octet FF at the
scale $2m_c$. This ad-hoc procedure brings monte carlo calculations into much better agreement with analytic NLL' calculations.
This suggests that if NRQCD fragmentation could be properly implemented in PYTHIA,  consistency with NLL' calculations would be obtained, though more work needs to be done on this problem. In Section~\ref{sec:conclusion} we give our conclusions. \app{evolve} summarizes the renormalization group evolution (RGE) needed for NLL' calculations
and also gives the profile functions that are used when computing the scale variation in the NLL' calculations. \app{match} describes the calculation of the matching coefficients and checks that they satisfy the required sum rules that relate them to the jet function.

\section{Fragmenting Jet Functions with Angularities}
\label{sec:FJFs}

In this section we extend the calculation of Ref.~\cite{Jain:2011xz} to FJFs with measured angularities.
We will follow the terminology of Ref.~\cite{Ellis:2010rwa}, in which a jet whose angularity is measured is referred to as a ``measured" jet,
while a jet for whom only the total energy is measured but
the angularity is not is called an ``unmeasured" jet. Here we consider the case of two particles as this is the most that
will appear in a one-loop calculation.
In Ref.~\cite{Jain:2011xz} the measurement operator in the definition of FJFs forces the mass squared of the jet to be $s$. The measurement
operator takes the form
\begin{equation}
\label{eq:eq1}
\delta(\omega(k^+-l^+-p^+))=\delta(s-\omega(l^+ + p^+)),
\end{equation}
where $k^{\mu}$ is the parent parton's momentum and $l^{\mu}$ and $p^{\mu}$ are the momenta of the partons carrying large lightcone components $l^- = (1-z) k^-$ and $p^- = z k^-$  of the parent's
momentum, respectively. The operator definition of the FJF with measured angularities is given by
\begin{align}
{\cal{G}}_i^h(\tau_a, z, \mu)= \int\frac{dk^{+} dp_h^+}{2 \pi} \int d^4y &\; e^{-i k^+ y^-/2} \\
& \times \sum_X \frac{1}{4N_C} \mathrm{tr} \Big{\lbrack}\frac{\nslash}{2} \langle 0 \vert \chi_{n,\omega}(y)\delta(\tau_a-\hat{\tau}_a)\vert Xh\rangle \langle Xh \vert \bar{\chi}_{n,\omega}(0)\vert 0\rangle \Big{\rbrack} \nn
\end{align}
where at $\cO(\as)$ the operator $\hat{\tau}_a$ takes the form (cf. \eq{tauadefn})
\begin{equation}
\label{eq:measop}
\delta(\tau_a-((l^+)^{1-a/2}(l^-)^{a/2}-(p^+)^{1-a/2}(p^-)^{a/2})/\omega)
\,.\end{equation}
Other than replacing \eq{eq1} with \eq{measop}, the integrals of all diagrams are the same
as in Ref.~\cite{Jain:2011xz}. However, rather than using the $\delta$-regulator and a gluon mass, we will use pure dimensional
regularization to regulate all divergences. In this limit, it is possible to show that the one-loop evaluation of the
FF yields
\begin{equation}
D_{i \rightarrow j}(z)= \delta_{ij} \delta(1-z) + T_{ij} \frac{\alpha_s}{2 \pi} P_{ij}(z) \left( \frac{1}{\epsilon_{UV}}-
\frac{1}{\epsilon_{IR}} \right)
\,,\end{equation}
where $T_{ij}$ are the color structures, $T_{qq}=C_{F},\; T_{gg}=C_{A},\; T_{qg}=C_{F},\; T_{gq}=T_{R}$.
Additionally, we have verified that the same $1/\epsilon_{IR}$ poles  appear in the calculation of FJFs and appropriately cancel
in the matching between the FJFs and FFs for all values of $a<1$. This justifies the formula
\begin{equation}
\label{eq:RepRule}
{\cal{G}}_i^h(\tau_a, z, \mu)=\sum_j \int_z^1 \frac{dx}{x}\; {\cal{J}}_{ij}(\tau_a, x, \mu) D_{j
\rightarrow h} \left( \frac{z}{x}, \mu \right),
\end{equation}
which is the analog of Eq.~(\ref{eq:ReplacementRule}) for FJFs that depend on the angularities.

Since the matching coefficients ${\cal{J}}_{ij}(\tau_a, z, \mu)$ are free of IR divergences, we can simplify the matching
calculation by using pure dimensional regularization, setting all scaleless integrals to zero and interpreting all
$1/\epsilon $ poles as UV. A detailed calculation of the renormalized finite terms of ${\cal{J}}_{ij}(\tau_a, z, \mu)$
can be found in \app{match}, the results of which are shown below. We parametrize the matching coefficients
${\cal{J}}_{ij}(\tau_a, z, \mu)$ as
\begin{align}
\begin{split}
\frac{{\cal{J}}_{ij}(\tau_a, z, \mu)}{2(2 \pi)^3} &= \delta_{ij}\delta(1-z) \delta(\tau_a) \\
&+ T_{ij} \frac{\alpha_s}{2 \pi } \Big{\lbrack}
c_0^{ij}(z,\mu)\delta(\tau_a) +c_1^{ij}(z,\mu) \left( \frac{1}{\tau_a}\right)_++c_2 \delta_{ij}\delta(1-z)
\left( \frac{\ln \tau_a}{\tau_a}\right)_+ \Big{\rbrack} \, ,
\end{split}
\end{align}
where
\begin{align}
\label{eq:CurlyJCoef}
c_0^{ij}(z, \mu) &= \frac{1-a/2}{1-a}\delta_{ij}\delta(1-z) \Big{\lbrack} \ln^2 \frac{\mu^2}{\omega^2}-\frac{\pi^2}{6}
\Big{\rbrack} + c^{ij}(z)\nn \\
&\quad - \bar{P}_{ji} \Bigg{[} \ln \frac{\mu^2}{\omega^2} + \frac{1}{1-a/2} \ln \left( 1+\left( \frac{1-z}{z} \right)^{1-a} \right)
+ (\delta_{ij}-1)\frac{1-a}{1-a/2}\ln (1-z) \Bigg{]}\, ,\nn \\
c_1^{ij}(z, \mu) &= -\frac{2}{1-a}\delta_{ij}\delta(1-z) \ln \frac{\mu^2}{\omega^2}+ \frac{1-a}{1-a/2} \bar{P}_{ij}\, , \nn\\
c_2 &= \frac{2}{(1-a)(1-a/2)} \, ,
\end{align}
with
\begin{align}
c^{qq}(z) &=1-z+\frac{1-a}{1-a/2}(1+z^2) \left( \frac{\ln(1-z) }{1-z} \right)_+ \nn \, ,\\
c^{gg}(z) &=\frac{1-a}{1-a/2}\frac{2(1-z+z^2)^2}{z} \left( \frac{\ln(1-z) }{1-z} \right)_+  \nn\, ,\\
c^{qg}(z) &=z  \nn\, ,\\
c^{gq}(z) &=2z(1-z)
\,,\end{align}
and where the $\bar{P}_{ij}$ are the splitting functions of Ref.~\cite{Jain:2011xz} except for the case $i=j=q$,
\begin{align}
\label{eq:split}
\begin{split}
&\bar{P}_{qq} = P_{qq}-\frac{3}{2}\delta(1-z)=\frac{1+z^2}{(1-z)_+}\, ,\\
&\bar{P}_{gg} = P_{gg} = 2\frac{(1-x+x^2)^2}{x(1-x)_+}\, ,\\
&\bar{P}_{qg} = P_{qg} = x^2 + (1-x)^2 \, ,\\
&\bar{P}_{gq} = P_{gq} = \frac{1+(1-x)^2}{x} \, .
\end{split}
\end{align}
Notice that our results for the matching coefficients ${\cal{J}}_{ij}(\tau_a,z,\mu)$ are independent of the jet algorithm and the jet size parameter $R$.
To include modifications of the ${\cal J}_{ij}(\tau_a,z,\mu)$
that come from these effects,  one would have to multiply the measurement operator in \eq{measop} by an additional $\Theta$-function
that imposes the phase space constraints required by the jet algorithm. However,  for jets with measured angularities, it was shown in Ref.~\cite{Ellis:2010rwa}
that jet-algorithm dependent terms for cone and $k_T$-type algorithms are suppressed by powers of $\tau_a/R^2$. Inuitively, this is because
as $\tau_a \to 0$ all the particles in the jet lie along the jet axis so the result must be insensitive to which algorithm is used and to the value of $R$ in this limit. For the values of $\tau_a$ and $R$ considered in this paper, $\tau_a/R^2$ is negligible and we will
drop these corrections.

As a non-trivial check of our results we show in \app{match} that our ${\cal{J}}_{ij}(\tau_a, z, \mu)$ satisfy
the following
identities and sum rules,
\begin{equation}
\lim_{a \rightarrow 0}{\cal{J}}_{ij}(\tau_a, z, \mu) =\omega^2 {\cal{J}}_{ij}(s, z, \mu) \, ,
\end{equation}
and
\begin{equation}
J_i(\tau_a, \mu)= \frac{1}{2(2 \pi)^3} \sum_j \int_0^1 dz \; z \; {\cal{J}}_{ij}(\tau_a, z, \mu)\, ,
\end{equation}
where ${\cal{J}}_{ij}(s, z, \mu)$ are the matching coefficients  for measured jet invariant mass found in Ref.~\cite{Jain:2011xz}
and $J_i(\tau_a, \mu)$ are the jet functions for measured jets that can be found in Ref.~\cite{Ellis:2010rwa}.

\section{$e^+e^- \to 2$ Jets with a $B$ Meson}
\label{sec:2jets}

In this section we present an analytic calculation of the cross section for $e^+e^-$ to two $b$ jets in which the $B$ meson is identified in a measured jet.
Following the analysis of Ref.~\cite{Ellis:2010rwa}, the factorization theorem for the cross section for one measured $b$ jet and one unmeasured $\bar{b}$ jet is
\begin{equation}
\label{eq:FacTh}
\frac{1}{\sigma_0} \frac{d \sigma}{d \tau_a} = H_2(\mu) \times S^{\mathrm{unmeas}}(\mu)\times J_{\bar{n}}^{(\bar{b})}(\mu)
\times \Big{\lbrack} S^{\mathrm{meas}}(\tau_a,\mu) \otimes J_{n}^{(b)} (\tau_a, \mu) \Big{\rbrack} ,
\end{equation}
where $H_2(\mu)$ is the hard function, $S^{\mathrm{unmeas}}(\mu)$ and $ S^{\mathrm{meas}}(\tau_a, \mu) $ are the unmeasured
and measured soft functions, $J_{\bar{n}}^{(\bar{b})}(\mu)$ is the unmeasured jet function containing the $\bar{b}$ quark
and $J_{n}^{(b)}( \tau_a, \mu)$ is the measured jet function containing the $b$ quark. These describe the short-distance process,
surrounding soft radiation, and radiation collinear to unmeasured and measured jets, respectively. At NLO the $\tau_a$-independent functions are given
by
\begin{align}
\begin{split}
H_2(\mu) &= 1 - \frac{\alpha_s(\mu) C_F}{2\pi} \left[ 8 - \frac{7\pi^2}{6} + \ln^2\frac{\mu^2}{\omega^2}
+ 3 \ln \frac{\mu^2}{\omega^2}\right]\, ,\\
S^{\text{unmeas}}(\mu) &= 1 + \frac{\alpha_s(\mu) C_F}{2\pi} \left[ \ln^2\frac{\mu^2}{4 \Lambda^2}
 - \ln^2 \frac{\mu^2}{4 \Lambda^2 r^2} - \frac{\pi^2}{3}\right] \, , \\
J_{\bar{n}}^{(\bar{b})}(\mu) &= 1 + \frac{\alpha_s(\mu) C_F}{2\pi}  J^q_{\rm alg} (\mu) ,
\end{split}
\end{align}
where $\Lambda$ is a veto on out-of-jet energy, $r =\tan{(R/2)}$ and
$J^q_{\rm alg}(\mu)$ is a function that depends on the algorithm used (and we will use the cone algorithm below) and is given in Eq.~(A.18) of Ref.~\cite{Ellis:2010rwa}.
We note that unlike measured jets, algorithm dependent contributions to the unmeasured jet are not power suppressed.
We also note that, beginning at $\cO(\alpha^2)$, non-global logarithms of the ratio $Q\tau_a/(2\Lambda r^2)$ begin to appear in the cross-section \cite{Chien:2015cka}. For the values of the parameters we consider, these ratios are such that we can treat these logarithms as $\cO(1)$ and thus these would enter as fixed order corrections needed at NNLL' accuracy, which is beyond the scope of this work.

We suppress the dependence of all these functions on scales other than the renormalization scale $\mu$.  Measured functions are convolved according to
\begin{equation}
f(\tau) \otimes g(\tau)  = \int d \tau^\prime \; f(\tau-\tau^\prime) g(\tau^\prime) .
\end{equation}
To calculate the differential cross section for a measured jet with an identified $B$ hadron, we apply the analogous replacement rule in Eq.~(\ref{eq:ReplacementRule})
to Eq.~(\ref{eq:FacTh}) and use the expression for the FJF in Eq.~(\ref{eq:RepRule}) to obtain
\begin{equation}
\label{eq:FacThB}
\frac{1}{\sigma_0} \frac{d \sigma^{(b)}}{d \tau_a dz} = H_2(\mu) \times S^{\mathrm{unmeas}}(\mu)\times J_{\bar{n}}^{(\bar{b})}(\mu)
\times  \sum_j  \Big{\lbrack} \Big{(}S^{\mathrm{meas}}(\tau_a, \mu) \otimes \frac{{\cal{J}}_{b j}^{(b)} (  \tau_a,z, \mu) }
{2 (2 \pi)^3} \Big{)} \bullet D_{j \rightarrow B}(z )\Big{\rbrack} ,
\end{equation}
where
\begin{equation}
G(z) \bullet F(z)=F(z) \bullet G(z) \equiv \int_z^1 \frac{dx}{x} F(x) G \Big{(}\frac{z}{x} \Big{)}.
\end{equation}
To obtain an NLL'  resummed formula for the cross section, we evaluate each function in the factorization theorem in Eq.~(\ref{eq:FacThB})
at its ``characteristic" scale (where potentially large logarithms are minimized) and, using renormalization group techniques, evolve
each function to a common scale, $\mu$, which we will choose to be equal to the hard scale. The details of this evolution are discussed in \app{evolve}.

The convolutions in Eq.~(\ref{eq:FacThB}) must be performed over angularity over $S^{\mathrm{meas}}$, ${\cal{J}}_{ij}$, and factors
arising from RG equations. Since such RG factors are distributions ($\delta$ or plus-distributions) in the angularity our final
answer is written in terms of distributions that can be computed analytically using Eqs.~(\ref{pd1}-\ref{pd3}).
Upon performing convolutions and resummation to NLL' accuracy we find for the cross section
\begin{align}
&d\sigma(\tau_a, z) \equiv \frac{1}{\sigma_0} \frac{d \sigma^{(b)}}{d \tau_a dz} = H_2(\mu_H) \times S^{\mathrm{unmeas}}
(\mu_{\Lambda})\times J_{\bar{n}}^{(\bar{b})}(\mu_{J_{\bar{n}}})  \times
\\
&\times  \sum_j  \Bigg{\{}\left( \frac{\Theta(\tau_a)}{\tau_a^{1+\Omega}} \right)
\Big{[}\delta_{bj} \delta(1-z) \left( 1+ f_{S}(\tau_a, \mu_{S^{\mathrm{meas}}})\right) + f_{{\cal{J}}}^{bj}(\tau_a, z, \mu_{J_n}) \Big{]}
\bullet \frac{D_{j \rightarrow B}(z, \mu_{J_{n}} )}{2 (2\pi)^3}  \nn\\
&\qquad \times \Pi (\mu, \mu_H, \mu_{\Lambda}, \mu_{J_{\bar{n}}}, \mu_{J_{n}},\mu_{S^{\mathrm{meas}}}) \ \Bigg{\}}_+ , \nn
\end{align}
where the `$+$' distribution is defined in \eq{plusdef} (and acts on all $\tau_a$-dependent quantities, including any implicit dependencies arising from the choice of scales $\mu_F$) and $\Omega(\mu_{J_{n}}, \mu_{S^{\mathrm{meas}}})= \omega_{J_n}(\mu, \mu_{J_{n}})+\omega_{S^{\mathrm{meas}}}(\mu_,
\mu_{S^{\mathrm{meas}}})$, the functions $ \omega_{J_n}$ and $\omega_{S^{\mathrm{meas}}}$ are given in \app{evolve}, the function $f_S$ is given by~\cite{Ellis:2010rwa}
\begin{align}
\begin{split}\label{eq:fS}
f_S(\tau,\mu)&= -\frac{\alpha_s(\mu) C_F}{\pi}\frac{1}{1-a}\left\{\left[ \ln \frac{\mu \tan^{1-a} \frac{R}{2}}{\omega \tau} +H(-1-\Omega)\right]^2 + \frac{\pi^2}{6} - \psi^{(1)}(-\Omega)\right\} \, ,
\end{split}
\end{align}
 and
$f_{\cal{J}}^{ij}$ are written in terms of the coefficients $c_0^{ij}$, $c_1^{ij}$ and $c_2$ presented in Eq.~(\ref{eq:CurlyJCoef}) as
\begin{align}
\begin{split}
f_{\cal{J}}^{ij}(\tau, z, \mu) &=  T_{ij} \frac{\alpha_s(\mu)}{2\pi} \Bigg( c_0^{ij}(z, \mu) + c_1^{ij} (z, \mu) \Big{(} \ln \tau -H(-1 -\Omega) \Big{)}   \\
& + c_2\delta_{ij} \delta (1-z)  \Big{(}\frac{(\ln \tau -H(-1-\Omega))^2+\pi^2/6-\psi^{(1)} (-\Omega)}{2} \Big{)} \Bigg).
\end{split}
\end{align}
The evolution kernel $\Pi$ is given in terms of $K_F(\mu, \mu_0)$ and $\omega_F (\mu, \mu_0)$ (cf.~\app{evolve}),
\begin{align}
&\Pi(\mu, \mu_H, \mu_{\Lambda}, \mu_{J_{\bar{n}}}, \mu_{J_{n}},\mu_{S^{\mathrm{meas}}}) = \prod_{F=H, J_{\bar{n}},
S^{\mathrm{unmeas}}} \exp ( K_F (\mu, \mu_F)) \left(  \frac{\mu_F}{m_F} \right)^{\omega_F (\mu, \mu_F)}
\\
&\qquad \times \frac{1}{\Gamma(-\Omega(\mu_{J_{n}}, \mu_{S^{\mathrm{meas}}}))} \times  \prod_{F= J_{n},S^{\mathrm{meas}}}
\exp ( K_F (\mu, \mu_F)+\gamma_E \omega_F (\mu, \mu_F)) \left(  \frac{\mu_F}{m_F} \right)^{j_F \omega_F (\mu, \mu_i)} \nn
\,,\end{align}
where $\mu_F, \; m_F$ and $j_F$ are given in Table~\ref{tb:scales}. Because they involve FFs (cf.~\app{match}), the $z$ convolutions must be evaluated numerically. For the fragmentation
of the $b$ quark we use a two-parameter power model FF introduced in Ref.~\cite{Kartvelishvili:1978jh}, in which
$D_{b\rightarrow B}(z,\; \mu = m_b = 4.5 \; \mathrm{GeV})$ is proportional to $z^{\alpha} (1-z)^{\beta}$.
Values for the parameters $\alpha=16.87$ and $\beta=2.628$ with $\chi^2_{d.o.f.}=1.495$ were determined using a fit to
LEP data in Ref.~\cite{Kniehl:2008zza} for the inclusive process $e^+e^- \rightarrow B+X$.
Errors in these parameters were not quoted in Ref.~\cite{Kniehl:2008zza}, so we cannot quantify errors associated with the extracted
FF in our calculation.
Additionally, we neglect the contribution from the fragmentation of other partons for our $e^+e^-$ collider studies as in Ref.~\cite{Kniehl:2008zza}. In proton-proton collisions at the LHC, gluon FJFs
must also be included since the dijet  channel  $gg\rightarrow gg$ gives a significant contribution to the production of jets with heavy flavor~\cite{Chien:2015ctp}.
For the evolution of the FF up to the jet scale we solve the DGLAP equation using an inverse Mellin transformation as
done in Ref.~\cite{Baumgart:2014upa}.

\begin{figure}[t]
\centerline{\includegraphics[scale=0.8]{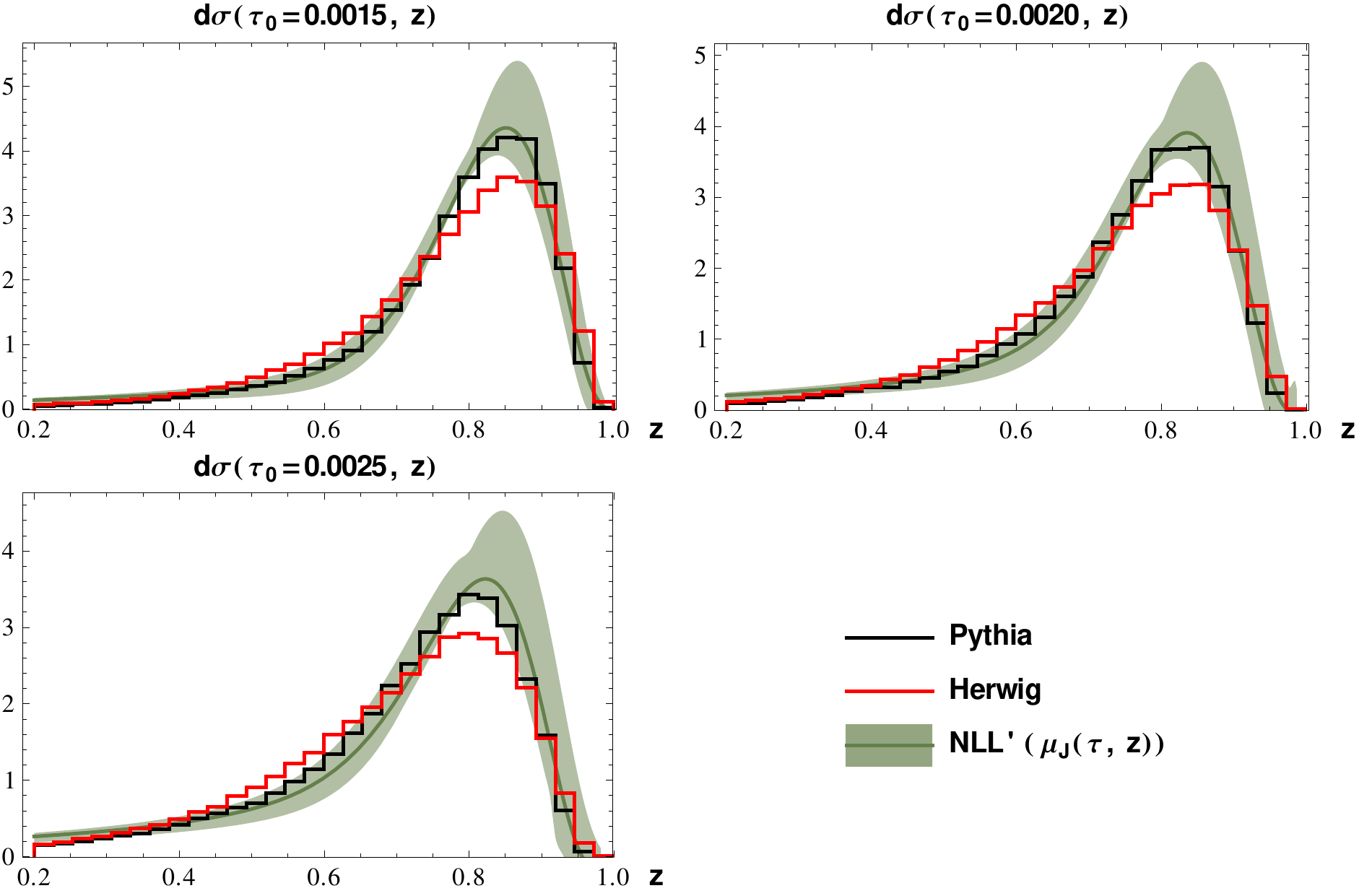}}
\caption{The $z$ distributions for $d \sigma(\tau_0, z)$ at $\tau_0= (1.5, \;2.0, \; 2.5) \times
10^{-3}$ for analytic calculations with theoretical uncertainty are shown in green. Monte carlo simulations using Madgraph $+$ PYTHIA
and Madgraph $+$ Herwig are shown in black and red, respectively.}
\label{fig:B_Var_tau}
\end{figure}

Fig.~\ref{fig:B_Var_tau} shows the $z$ distributions from $d\sigma(\tau_0, z)$ for $\tau_0= (1.5, \;2.0, \; 2.5 )\times
10^{-3}$ of our analytic NLL' calculation (green) and  monte-carlo simulations using Madgraph $+$ PYTHIA (black)
and Madgraph $+$ HERWIG (red). For each monte carlo and for each NLL' calculation,
the graphs are independently normalized to unit area. For plots with fixed $\tau_a$ we use a $z$-bin of $\pm \,0.1$ and
for plots with fixed $z$ we use a $\tau_a$ bin of size $\pm \, 2\times 10^{-4}$. Jets are reconstructed in PYTHIA using the
Seedless-Infared-Safe Cone (SISCONE) algorithm in the FastJets package~\cite{Cacciari:2011ma} with $R=0.6$, which will be used throughout this work.
We produced simulated dijet events at $E_{cm}=250$ GeV in which each jet has an energy of at least $(E_{cm}-\Lambda)/2$ where $\Lambda = 30$ GeV.\footnote{This is different than simply placing a cut $\Lambda$ on energy outside the jets (which is what is assumed in our analytical results), but this difference only appears at $O(\alpha_s^2)$ in the soft function,
which is higher order than we work in this paper.}
\begin{figure}[t]
\centerline{\includegraphics[scale=0.8]{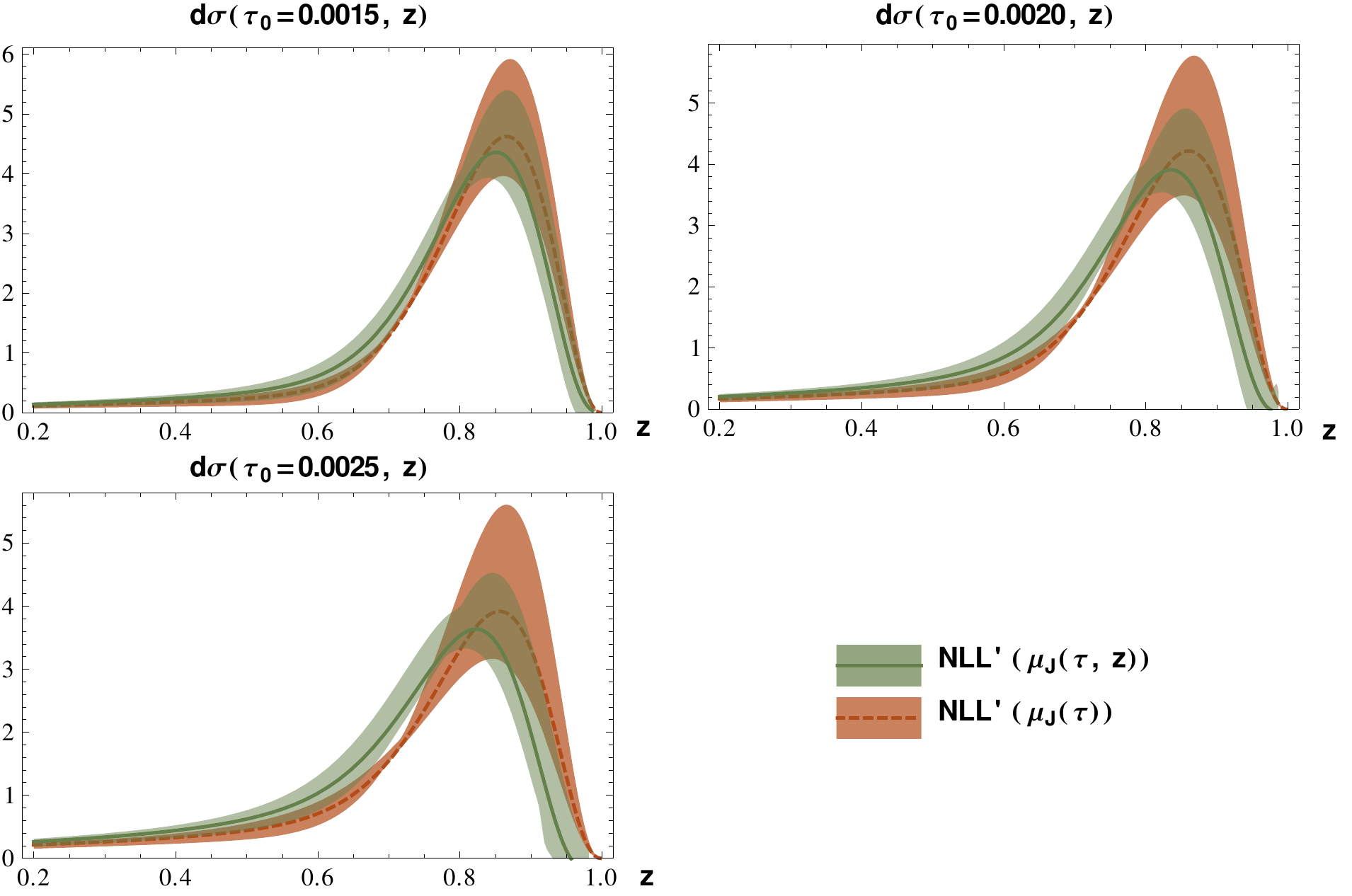}}
\caption{Analytic results for the z distributions of $d \sigma(\tau_0, z)$ at $\tau_0= (1.5, \;2.0, \; 2.5)\times
10^{-3}$. The orange curve  is calculated with a measured jet scale that does not depend on $z$
whereas the green curve uses a scale that does depend on $z$ (as in Fig.~\ref{fig:B_Var_tau}).}
\label{fig:B_z_vs_no_z}
\end{figure}
The central green line corresponds to the NLL' calculation with the various functions in the factorization
theorem evaluated at their characteristic values shown in Table~\ref{tb:scales}, and the green band corresponds to the
estimate of theoretical uncertainty obtained by varying the scales of the unmeasured
functions by $\pm 50 \%$, and  using profile functions
~\cite{Ligeti:2008ac, Abbate:2010xh, Hornig:2016ahz} to estimate the
uncertainty of the measured functions. Profile functions allow us to introduce an angularity dependent scale variation that freezes at the
characteristic scale for high values of $\tau_a$ where the factorization theorem breaks down and at a fixed scale for small values
of $\tau_a$ where we reach the non-perturbative regime. This method for estimating theoretical uncertainties is used throughout this work.
Additional details on the profile functions we use can be found in \app{evolve}.

\begin{table}[h]
\begin{center}
\begin{tabular}{|c||c|c|c|c|c|}
\hline
Function $(F)$ & $H_2$ &  $J_{\bar{n}}^{\bar{b}}$ & $S^{\mathrm{unmeas}}$ & ${\cal{J}}(\tau, z)$ & $S^{\mathrm{meas}}(\tau)$\\
\hline
\hline
Scale $(\mu_F)$ & $E_{\mathrm{cm}}$ & $\omega_{\bar{n}} r$ &$2 \Lambda r^{1/2}$& $\omega_{n}\tau^{1/(2-a)}(1-z)^{(1-a)/(2-a)}$&
$\omega_{n}\tau/r^{1-a}$\\
\hline
$m_F$ & $\omega$ & $w_{\bar{n}} r$ & $2 \Lambda r^{1/2}$& $\omega_n$&$\omega_{n}/r^{1-a}$\\
\hline
$j_F$& $1$& $1$& $1$& $2-a$& $1$\\
\hline
\end{tabular}
\end{center}
\caption{Characteristic scales of the different functions in the factorization theorem of Eq.~(\ref{eq:FacTh}). }
\label{tb:scales}
\end{table}

The orange curves in Fig.~\ref{fig:B_z_vs_no_z} show the differential cross section as a function of $z$ for fixed $\tau_0$ where $\mu_J(\tau) = \mu_J (\tau, z=0)=\omega \tau^{1/(2-a)}$
is chosen as the characteristic scale of the measured jet function, and the error band is obtained the same way as for Fig.~\ref{fig:B_Var_tau}. As in Fig.~\ref{fig:B_Var_tau}, the green curves show the
cross section for a measured jet scale $\mu_J (\tau, z)=\omega \tau^{1/(2-a)}(1-z)^{(1-a)/(2-a)}$.
\begin{figure}[t]
\centerline{\includegraphics[scale=0.8]{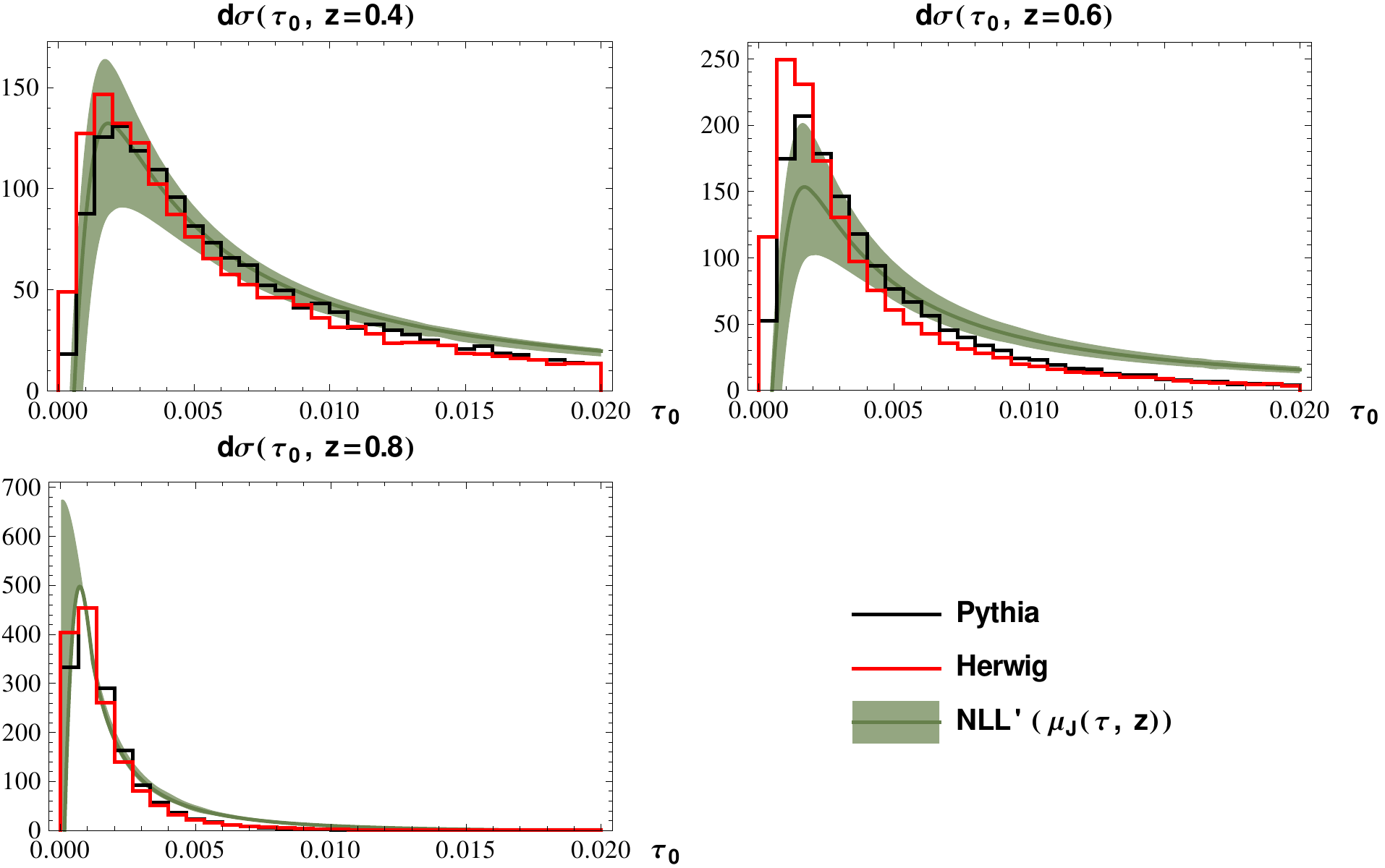}}
\caption{Angularity distributions of $d\sigma(\tau_a, z)$ for $a=0$ at $z=0.4,\;0.6,\;0.8$. Analytic results are
shown as green bands. Monte carlo results are shown as black lines for Madgraph $+$ PYTHIA and red lines for Madgraph $+$ HERWIG.}
\label{fig:B_Var_z}
\end{figure}
The reorganization of logarithms of $(1-z)$ shown in Eq.~(\ref{fconf}) suggests that we can improve the accuracy of our calculations for
$z \rightarrow 1$ by choosing the characteristic value of the measured jet scale to be $\mu_J (\tau, z)$. This improvement is clearly seen
in Fig.~\ref{fig:B_z_vs_no_z} which shows the scale variation for the choices $\mu_J(\tau)$ and $\mu_J(\tau,z)$, the latter choice gives smaller scale variation near the peak in the $z$ distribution.

In Fig.~\ref{fig:B_Var_z} we present the results for the $\tau_0$ distributions of the differential cross section $d\sigma (\tau_a,z)$
for $z=0.4,\; 0.6,\; \mathrm{and}\; 0.8$. The color and normalization schemes match those in Fig.
\ref{fig:B_Var_tau}.
We see that for higher values of $z$ the distributions of
$\tau_0$ are shifted towards smaller values. This is expected since the majority of the energy of the jet is carried by the B meson
which results in narrower jets. Figs.~\ref{fig:B_Var_tau} and~\ref{fig:B_Var_z} show that our results are consistent within the
monte carlo uncertainty that is suggested by the difference between PYTHIA and HERWIG predictions. This gives us confidence that the FJF formalism combined with NLL' resummation can be used to correctly calculate both the
substructure and the identified hadron's energy fraction within a jet.
\section{$e^+e^- \to 3$ Jets with the Gluon Jet Fragmenting to $ J/\psi$}
\label{sec:3jets}
We can also use the FJF formalism to calculate the cross section for $e^+e^- \to 3$ jets with a $J/\psi$.
As we expect gluon fragmentation to be the dominant production channel at the LHC, we focus
on the case where $J/\psi$ is found within a gluon jet. In addition, we assume that the angularity of this jet is also measured. To
obtain a physical observable, one must also include contributions from all jets  fragmenting to  $J/\psi$, however, we expect
the contribution from quark jets to be smaller.  It is theoretically possible to isolate the $J/\psi$ coming from gluon jets in experiments by $b$-tagging
the other two jets in the event,
so we will focus on the process $e^+e^- \to b\bar{b} g$ followed by gluon fragmentation to $J/\psi$.

The analytic  expression for this cross section is
\begin{align}
&\frac{1}{\sigma_0} \frac{d \sigma^{(g)}}{d \tau_a dz} = H_3(\mu_H) \times S^{\mathrm{unmeas}}(\mu_{\Lambda})
\times J_{n_1}^{(\bar{b})}(\mu_{J_{n_1}}) \times J_{n_2}^{(b)}(\mu_{J_{n_2}})
\nn\\
&\times \sum_i \!\Bigg{\{}\!\left( \frac{\Theta(\tau_a)}{\tau_a^{1+\Omega}} \right) \!\Big{[}\delta_{gi}
\delta(1-z) ( 1+ f_{S}(\tau_a, \mu_{S^{\mathrm{meas}}})) + f_{{\cal{J}}}^{gi}(\tau_a, z, \mu_{J_{n_3}}) \Big{]}
\bullet \frac{D_{i \rightarrow J/\psi}(z,\mu_{J_{n_3}} )}{2(2\pi)^3}
\nn\\
&\qquad \times \Pi (\mu, \mu_H, \mu_{\Lambda}, \mu_{J_{n_1}}, \mu_{J_{n_2}},\mu_{J_{n_3}}, \mu_{S^{\text{meas}}})\! \Bigg{\}}_+
\,,\end{align}
where $\Omega\equiv \Omega(\mu_{J_{n_3}}, \mu_{S^{\mathrm{meas}}})= \omega_{J_n}(\mu, \mu_{J_{n_3}})+\omega_{S^{\mathrm{meas}}}(\mu_,
\mu_{S^{\mathrm{meas}}})$, the $b$-quark initiated jets $J_{n_1}^{(b)}$ and $J_{n_2}^{(\bar{b})}$ are unmeasured, the expression for $f_S$ is the same as Eq.~(\ref{eq:fS}) with $C_F$ replaced by $C_A$,  and our expressions for $f_{\cal{J}}^{ij}$ are  given in terms of the coefficients $c_0^{ij}$, $c_1^{ij}$ and $c_2$
given in \eq{CurlyJCoef}. Here $\sigma_0$ is the LO cross section for $e^+e^-\to b \bar{b} g$. We will focus on the Mercedes Benz configuration in which all three jets
have (approximately) the same energy, and consider jets with energies large enough that the mass of $b$-quark can be neglected. Here, $H_3(\mu)$ is $1 +O(\alpha_s)$ where
the $O(\alpha_s)$ comes from the NLO virtual corrections to $e^+e^- \to b\bar{b} g$. We do not include this correction. The primary effect of its omission will be on the normalization of the cross section, which is not important for our discussion of the distributions we show below, and to increase the scale uncertainty associated with varying $\mu_H$; however this is not a very important source of uncertainty in our calculations.

While the calculation for $B$ mesons requires a phenomenological FF, the FFs for $J/\psi$ production
can be calculated in NRQCD~\cite{Bodwin:1994jh}.
Refs.~\cite{Braaten:1993mp,Braaten:1993rw,Braaten:1994vv,Braaten:1996jt} showed that a
$J/\psi$ FF can be calculated in terms of analytically calculable functions of $\alpha_s(2m_c)$ and $z$
multiplied by nonperturbative NRQCD  long-distance matrix-elements (LDMEs). In $J/\psi$ production, the most important  production mechanisms
are the color-singlet mechanism, in which the $c\bar{c}$ is produced perturbatively  in a $^3S_1^{(1)}$ state, and the color-octet mechanisms,
in which the  $c\bar{c}$ is produced perturbatively  in a $^1S_0^{(8)}$, $^3S_1^{(8)}$, or $^3P_J^{(8)}$ state.
Here $^{2S+1}L_J^{(1,8)}$ refers to the angular momentum and color quantum numbers of the $c\bar{c}$.
The numerical values for the corresponding  LDMEs are taken to be the central values from the global fits performed in  Refs.~\cite{Butenschoen:2011yh, Butenschoen:2012qr},
and are shown in Table \ref{tb:LDME}.
\begin{table}[h]
\begin{center}
\begin{tabular}{|l|c|c|c|c|c|}
\hline
$\langle\mathcal{O}^{J/\psi}(^3S_1^{(1)}) \rangle$  & $\langle \mathcal{O}^{J/\psi}(^3S_1^{(8)}) \rangle $ &
  $\langle \mathcal{O}^{J/\psi}(^1S_0^{(8)}) \rangle$ & $\langle\mathcal{O}^{J/\psi}(^3P_J^{(8)})\rangle /m_c^2$ \\
\hline
1.32 GeV$^3$& 2.24 $\times 10^{-3}$ GeV$^3$ & $4.97 \times 10^{-2}$ GeV$^3$ & -7.16 $\times 10^{-3}$ GeV$^3$\\
\hline
\end{tabular}
\end{center}
\caption{LDMEs for NRQCD production mechanisms. We use central values taken from global fits in Refs.~\cite{Butenschoen:2011yh, Butenschoen:2012qr}. }
\label{tb:LDME}
\end{table}
The color-singlet LDME scales as $v^3$, where $v$ is the typical relative velocity of the $c\bar{c}$ in the $J/\psi$,
while the color-octet LDMEs scale as $v^7$~\cite{Bodwin:1994jh}. This $v^4$ suppression is clearly seen in the numerical values of the LDMEs in Table \ref{tb:LDME}.
In the calculation of the gluon FF, this $v$ suppression is compensated by powers of $\alpha_s$ since the leading color-octet contributions are $O(\alpha_s^2)$ in the
$^1S_0^{(8)}$ and $^3P_J^{(8)}$ channels and $O(\alpha_s)$ in the $^3S_1^{(8)}$ channel, while the color-singlet contribution is $O(\alpha_s^3)$.
In this work we focus on the gluon FJF, $\mathcal{G}_g^{J/\psi}$, and separately compute each of the four NRQCD contributions
to $\mathcal{G}_g^{J/\psi}$. To calculate $\mathcal{G}_g^{J/\psi}$, we evolve each
FF from the scale $\mu=2m_c$ to the characteristic scale of the measured jet $\mu_{J_{n_3}}(\tau_a)=\omega \tau_a^{1/(2-a)}$
using the DGLAP evolution equations. For most values of $z$ considered in this section, we do not expect that using a $z$ dependent
scale will result in significant improvement in the scale variation. In addition, using a $z$ dependent scale in the $^3P_J^{(8)}$ channel
yields unphysical results, such as negative values for the FF.  After evolution, we perform the convolution $\left[D \bullet f_{\cal{J}}\right](z)$
in $z$ with the matching coefficients derived in Section~\ref{sec:2jets}.

Before discussing the comparison of our results with monte carlo, we briefly review how the Madgraph $+$ PYTHIA monte carlo handles color-singlet and color-octet quarkonium production.
We produce quarkonia states in Madgraph from the following processes: $e^+e^- \to b\bar{b}ggc\bar{c}[^3S_1^{(1)}],\; e^+e^- \to b\bar{b}gc\bar{c}[^1S_0^{(8)}]$, and $e^+e^- \to b\bar{b}c\bar{c}[^3S_1^{(8)}]$. The quantum numbers $^{2S+1}L_J^{(1,8)}$ are for the $c\bar{c}$ produced in the event. We only include diagrams in which the virtual photon couples to the $b\bar{b}$ so in all cases the $c\bar{c}$ plus any additional gluons come from the decay of a virtual gluon.
We did not simulate production in the $^3P_J^{(8)}$ channel in $e^+e^-\to b\bar{b} g \to b\bar{b} c\bar{c}g$ because IR divergences in the matrix elements require much longer running times to get the same number of events. We then
perform showering and hadronization on these hard processes using PYTHIA. Analysis is done using RIVET~\cite{Buckley:2010ar}.
During PYTHIA's showering phase, color-singlet $J/\psi$ do not radiate gluons. Thus if these $J/\psi$ are produced within a jet, all surrounding radiation
is due to the other colored particles in the event~\cite{Sjostrand:2006za, Sjostrand:2014zea}. We require that after showering there are only three jets in the event, two from the
$b$-quarks and one from a gluon that contains the $J/\psi$.
We simulate three-jet events
at $E_{cm}=250$ GeV in the Mercedes-Benz configuration by requiring the jets each have energies $E_{jet} > (E_{cm}-\Lambda)/3$ with $\Lambda=30$ GeV, analagous to what was done in \sec{2jets}.

For $c\bar{c}$ produced in a color-octet state PYTHIA allows the color-octet $c\bar{c}$ to emit gluons with a splitting function $2 P_{qq}(z)$.
Since $P_{qq}(z)$ is peaked at $z=1$, the color-octet $c\bar{c}$ pair typically retains most of its energy after these emissions. This model of the production mechanism is very different than the physical process implied by the NLL' calculation. In the NLL' calculation, the FF is calculated at the scale $2m_c$, then evolved up to the jet energy scale using Altarelli-Parisi evolution equations. Since this is a gluon FF, the most important splitting kernel in this evolution is $P_{gg}(z)$. We find that the FFs obtained at the jet energy scale are not significantly changed if we use only this evolution kernel and ignore mixing with quarks. Thus the production process implied by the NLL' calculation is that of a highly energetic gluon produced in the hard process with virtuallity of order the jet energy scale, which then showers by emitting gluons until one of the gluons with virtuality of order $2m_c$ hadronizes into the $J/\psi$. Because
$P_{gg}(z)$ is peaked at $z=0$ and $z=1$ the resulting $J/\psi$ distribution in $z$ is much softer than the model employed by PYTHIA.
PYTHIA does not allow one to change the actual splitting function, only to modify the color-factor. Therefore, in order to get a softer $z$ distribution we changed the coefficient of PYTHIA's splitting kernel for a gluon radiating off a color-octet $c\bar{c}$ pair from $2 P_{qq}$ to $C_A P_{qq}=3 P_{qq}$.
This results in a slighter softer $z$ distribution than default PYTHIA, but is still inconsistent with the NLL' calculation. This change does not have significant impact on the $\tau_a$ distributions.
The $\tau_a$ distributions are generally in better agreement. The variable $\tau_a$ depends on all of the hadrons in the jet and is therefore
less sensitive to the behavior of the $J/\psi$, especially when the $J/\psi$ carries a small fraction of the jet energy. In that case, $\tau_a$ distributions in the NLL' calculation
look similar for all color-octet mechanisms.

In an attempt to see if PYTHIA can be modified to reproduce the $z$ distributions obtained in our NLL' calculations,
and confirm the physical picture of the NLL' calculation described above,  we generate $e^+e^- \to b\bar{b}g$ events in Madgraph and allow PYTHIA to shower but not hadronize
the events. If we allow the shower to evolve to a scale where the typical invariant mass of a gluon is $2 m_c$ and then convolve the gluon distribution with the NRQCD FFs at this scale, we expect that  the resulting $z$ distributions should mimic our NLL' calculation.
The lower cutoff scale in PYTHIA's parton shower is set by the parameter TimeShower:pTmin, which is related to the minimal virtuality of the particles in the shower, and whose default value is 0.4 GeV. We change this parameter to 1.6 GeV, which corresponds to a virtuality of $\sim 2m_c$, then obtain a $z$ distribution for the gluons by randomly choosing a gluon from the gluon initiated jet. We then numerically convolve this $z$ distribution with the analytic expression for the NRQCD FF. This procedure, which we will refer to  as Gluon Fragmentation Improved PYTHIA (GFIP), yields $z$ distributions that are consistent with our NLL' result, as we will see below.  We tested an analogous procedure for two-jet events with $B$ mesons by showering
$e^+e^-\to b\bar{b}$ with PYTHIA with hadronization turned off. We then
convolved the resulting $b$ quark distribution with the  $b$-quark FF at
the scale $2 m_b$, and found results for $B$ mesons that are consistent with our NLL' calculations. Note that PYTHIA treats the radiation coming from the octet $c\bar{c}$ pair the same regardless of the angular momentum quantum numbers. In contrast,  GFIP  like the NLL' calculation gives different results for all three channels by applying different FFs at the end of the parton shower phase. Also GFIP can be applied to all four NRQCD production mechanisms, since
convergence issues for the $^3P_J^{(8)}$ channels are absent.

\begin{figure}[t]
\centerline{\includegraphics[scale=0.75]{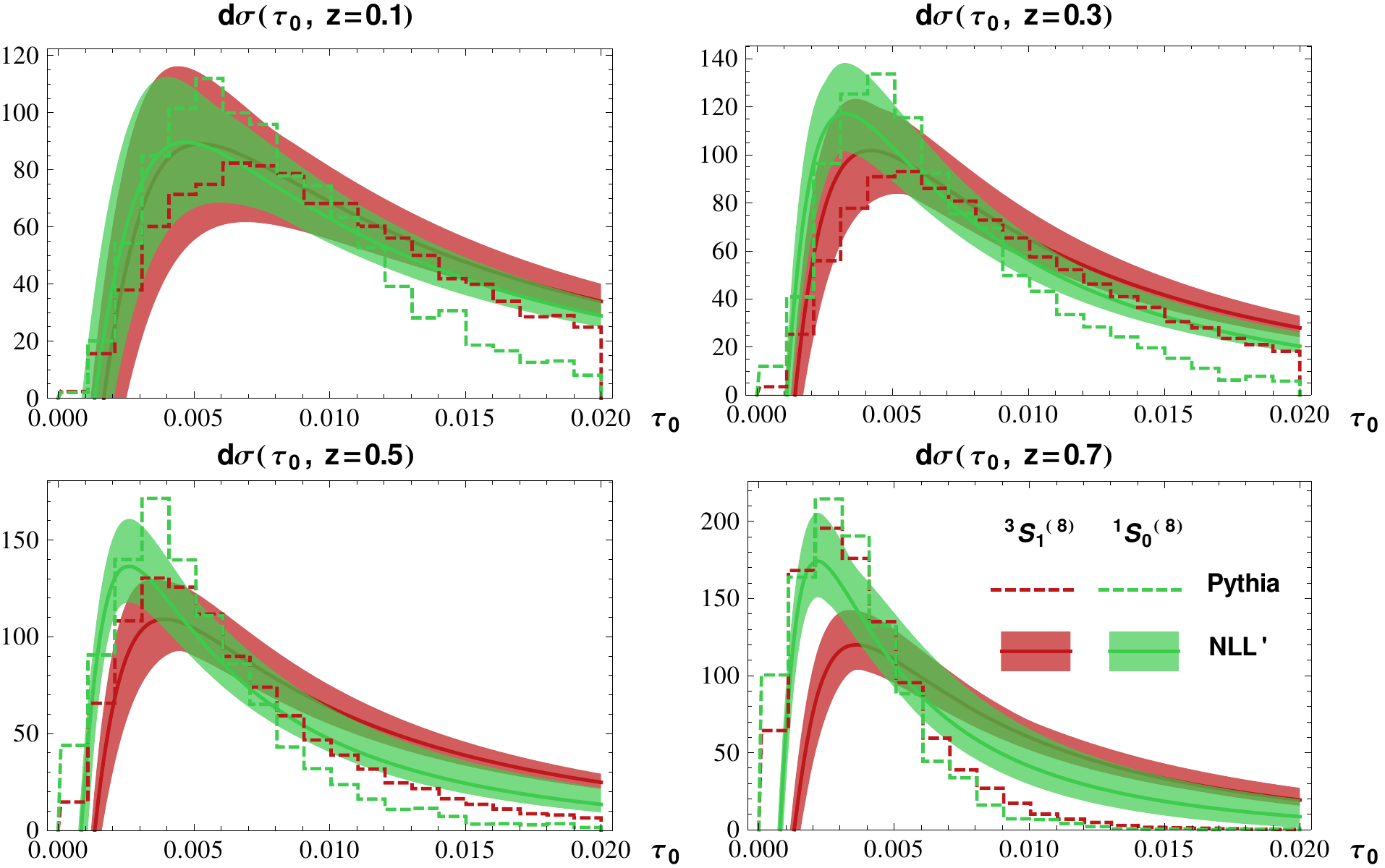}}
\caption{Angularity distributions of $d\sigma(\tau_a, z)$ for $a=0$ at $z=0,1,\; 0.3,\;0.5,\;0.7$. Analytic calculations are shown
as red (green) bands for the $^3S_1^{(8)}$ ($^1S_0^{(8)}$)
production mechanisms. Results from Madgraph $+$ PYTHIA are shown as red (green) dashed lines for the same mechanisms.}
\label{fig:Jpsi_Var_z}
\end{figure}

\begin{figure}[t]
\centerline{\includegraphics[scale=0.75]{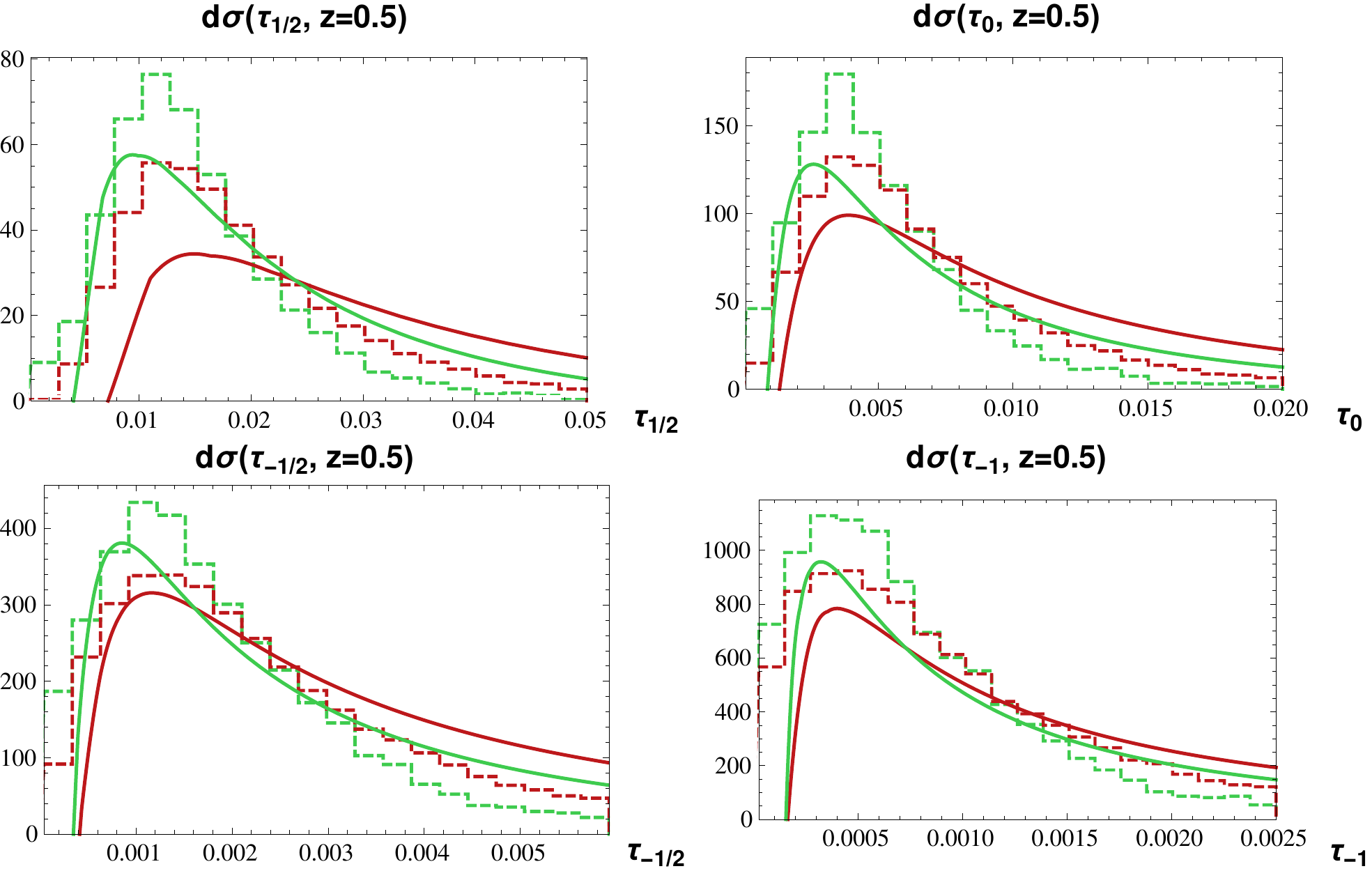}}
\caption{Angularity distributions of $d\sigma(\tau_a, z)$ for $a=+1/2,\,0,\,-1/2,\,-1$ at $z=0.5$. Analytic calculations are shown
as red (green) solid lines for the $^3S_1^{(8)}$ ($^1S_0^{(8)}$)
production mechanisms. Results from Madgraph $+$ PYTHIA are shown as red (green) dashed lines for the same mechanisms.}
\label{fig:Jpsi_Var_a}
\end{figure}

Fig.~\ref{fig:Jpsi_Var_z} shows our NLL' calculation and Madgraph $+$ PYTHIA results for the distribution of $\tau_0$ for various
fixed values of $z$ for the
$^3S_1^{(8)}$ (red) and $^1S_0^{(8)}$  (green) channels. We see fairly good agreement between analytic and Monte Carlo results in the peak regions
for smaller values of $z$ and notice some qualitative differences in the tail regions, especially for the $^1S_0^{(8)}$ channel. At higher
values of $z$ where the number of final state particles is small, differences in the $\tau_0$ distributions could be attributed to the
increasing influence of Pythia's unrealistic model of quarkonium production. As $z \to0$, we also see similar $\tau_0$ dependence for the
two color-octet channels in our analytic results. This suggests that in the small $z$ region, the jet substructure is independent
of the production mechanism. Thus, attempts to use angularity distributions to extract the various LDMEs should focus on the range
$0.3 < z < 0.7$.


In Fig.~\ref{fig:Jpsi_Var_a}, we show the angularity distributions (without uncertainties) for the $^1S_0^{(8)}$ and $^3S_1^{(8)}$
mechanisms for
$a=+1/2,\,0,\,-1/2,\,-1$. These are computed analytically and using monte carlo and we again see reasonable agreement. As $a$ is decreased,
we see less discrimination
between the two production mechanisms. Thus extraction of LDMEs should ideally be done with larger values of $a$, for $a < 1$ where
factorization in $\mathrm{SCET}_I$ holds, with the caveat that there is a trade-off since the predictability of the analytical results
is limited for $a$ too close to $1$ since power corrections grow as $1/(1-a)$ \cite{Lee:2006nr}.
\begin{figure}[t]
\centerline{\includegraphics[scale=0.5]{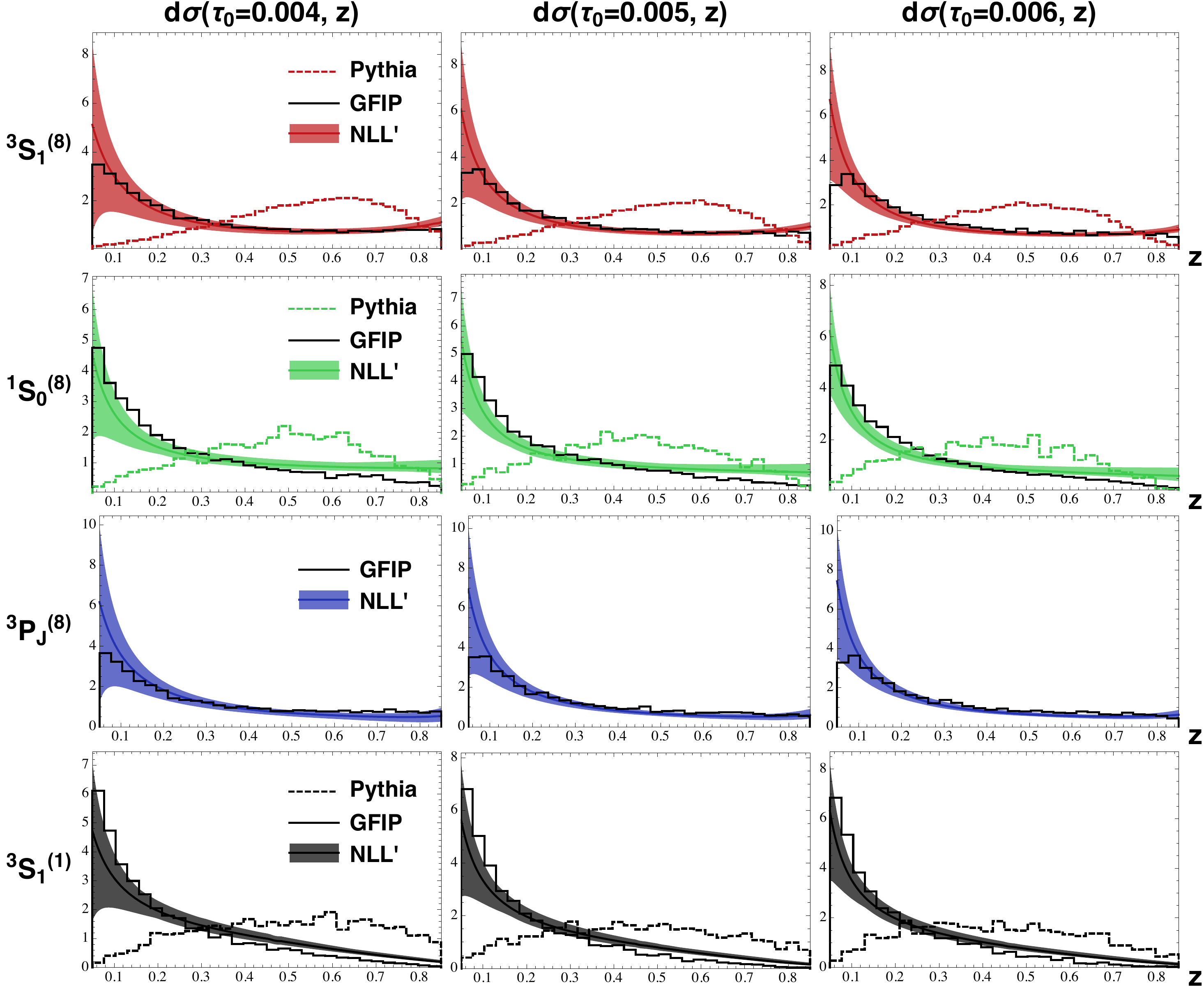}}
\caption{$z$ distributions of $d\sigma(\tau_a, z)$ for NLL' analytic calculations (bands), PYTHIA (dashed lines), and GFIP (solid lines) for fixed
values of $\tau_0=(4,\;5,\;6)\times10^{-3}$.}
\label{fig:Jpsi_z_All}
\end{figure}

In contrast to the angularity distributions, Fig.~\ref{fig:Jpsi_z_All} shows that analytic and monte carlo calculations of the $z$ distributions
using Madgraph $+$ PYTHIA yield strikingly different results, with Madgraph+PYTHIA yielding a much harder $z$-distribution. Fig.~\ref{fig:Jpsi_z_All} also shows the $z$ distributions using GFIP. The GFIP modification yields significantly different results for the $z$ distributions that align
more closely with NLL' calculation. While this is far from a proper modification of PYTHIA, it shows us that
implementing the missing $g \to J/\psi$ fragmentation yields encouraging similarities to our analytical calculations using the FJF formalism
with NRQCD FFs. This also suggests that if monte carlo is modified to properly include NRQCD FFs at the scale $2m_c$ it will yield results that are consistent with
FJFs combined with NLL' resummation. Correct monte carlo implementation of the NRQCD FFs is important because the GFIP modification can only be used to  calculate the $z$ distribution. There are many other jet shape observables,  such as $N$-subjettiness or $\Delta R$ (where $\Delta R$ is the angle between the $J/\psi$ and the jet axis), that should
be able to discriminate between NRQCD production mechanisms, and many of these are most easily predicted using monte carlo.

\section{Conclusion}
\label{sec:conclusion}
The study of hadrons within jets provides new tests of perturbative QCD dynamics. The distribution in  $z$ (the fraction of jet energy carried by the identified hadron)
can be calculated as a convolution of the well-known fragmentation functions (FFs) for that hadron with perturbative matching coefficients that are calculable at the jet energy scale,
which is typically well above $\Lambda_{\rm QCD}$. At hadron colliders this provides a new way to extract FFs and will be especially important for pinning down gluon FFs, which are of
subleading importance in $e^+e^-$ colliders where FFs are usually measured.  The production of heavy quarkonia within high energy jets in collider experiments also provides new tests of NRQCD.

In this paper, we studied cross sections for jets with heavy mesons as a function of  $z$ and the substructure variable angularity, $\tau_a$. We provided for the first time the NLO
matching coefficients for jets with measured $\tau_a$, and used these along with the known RGE  for the hard, jet, and soft functions to obtain
NLL' accuracy calculations of cross sections for jets with heavy mesons. We considered the production of $B$ mesons in two-jet events in $e^+e^-$ collisions at $E_{\rm cm}=250$
GeV as well as $J/\psi$ production in three-jet events at the same energies. Though not relevant to any experiment,  this is useful for comparing NLL' calculations
with monte carlo simulations of fragmenting jets whose energy is comparable to those measured at the LHC. In the simulations of quarkonia production, the underlying hard process was generated using Madgraph and then PYTHIA was used to shower and hadronize the events. In the simulations involving B meson production we also used HERWIG.

For $B$ mesons,  we find that the $z$ and $\tau_a$ distributions computed using monte carlo and NLL' are in excellent agreement, giving us confidence in our analytic approach.
In the case of $J/\psi$, we considered three-jet events in which the jets all had the same energy and the $J/\psi$ in both simulation and NLL' calculations was required to
come from the gluon jet. This allowed us to study $J/\psi$ production via the fragmentation of high energy  gluon initiated jets, which we expect to be  an important mechanism
at the LHC. Earlier studies of gluon FJFs in Ref.~\cite{Baumgart:2014upa} indicated that the $z$ and $E$ dependence of these jets could discriminate between various NRQCD
production mechanisms. The analytic NLL' studies of this paper are consistent with Ref.~\cite{Baumgart:2014upa}; we also find that the $\tau_a$ and $z$ distributions can
discriminate between different various NRQCD production mechanisms.

For monte carlo simulations, we used Madgraph to calculate $e^+e^-\to b\bar{b} g$ followed by the gluon fragmenting into a a $c\bar{c}$ pair in either a  $^3S_1^{(8)}$, $^1S_0^{(8)}$, or $^3S_1^{(1)}$ state. As explained earlier we do not simulate events in the $^3P_J^{(8)}$ channel.
 The events were then showered  and hadronized using PYTHIA.
While the $\tau_a$ distributions are similar to analytical calculations, the $z$ distributions are much harder and their shape looks nothing like the NLL' calculation.
We attribute this to a naive model that PYTHIA uses for simulating the radiation of gluons from color-octet $c\bar{c}$ pairs.

We then considered an alternative simulation approach where $e^+e^- \to b\bar{b}g$ events are generated using Madgraph, then PYTHIA is used to shower  the event
to a low scale near $2 m_c$ without hadronization. The resulting gluon distribution is then convolved with the analytically calculated NRQCD FFs calculated at the scale $2 m_c$.
This procedure yields $z$ distributions that are in much better agreement with our NLL' calculations.

Future work will focus on extending the NLL' calculations to hadron colliders, where the unmeasured jet and soft function recently calculated in Ref.~\cite{Hornig:2016ahz} must
be combined with the FJFs of this paper. It would be of great interest to compare the results of these calculations with data from the LHC on high energy jets with heavy mesons
and quarkonia. Finally, there needs to be more work on improving the understanding of the differences between NLL' and monte carlo simulations. Monte carlo simulations that can
properly simulate the production of quarkonia within jets will be essential for calculating other jet observables  for which NLL' calculations are either unavailable or impractical.
\acknowledgments
AH was supported by a Director's Fellowship from the  LANL/LDRD program and the DOE Office of Science under
Contract DE-AC52-06NA25396. TM and YM are supported in part by the Director, Office of Science, Office of
Nuclear Physics, of the U.S. Department of Energy under grant numbers DE-FG02-05ER41368. RB is supported by
a National Science Foundation Graduate Research Fellowship under Grant No. 3380012. RB, TM, and YM also
acknowledge the hospitality of the theory groups at Brookhaven National Laboratory, Los Alamos National Laboratory,
Duke-Kunshan University, and UC-Irvine for their hospitality during the completion of this work.  AL and LD were supported
in part by NSF grant PHY-1519175.  LD also acknowledges the hospitality of the theory group at Duke University
for their hospitality during the completion of this work.

\appendix

\section{Renormalization Group and Resummation}
\label{evolve}

\subsection{Evolution of Measured and Unmeasured Functions}

The RGEs satisfied by the elements of the factorization theorem are separated into two categories; terms that do depend on the
variable $\tau_a$ and terms that do not. The latter satisfy the following RGE
\begin{equation}
\mu \frac{d}{d \mu} f(\mu) = \gamma_f (\mu) f(\mu) \, ,
\end{equation}
where $ \gamma_F (\mu)$ is the anomalous dimension
\begin{equation}
\gamma_F(\mu) = - \frac{1}{Z_F(\mu)} \mu \frac{d}{d \mu} Z_F(\mu) = \Gamma_F (\alpha_s) \ln \left( \frac{\mu^2}{m_F^2} \right) +
\gamma_F(\alpha_s) \, ,
\end{equation}
where $m_F$ is related to the characteristic scale for the particular function,
and $Z_F (\mu)$ is the renormalization function for $F(\mu)$. The coefficient $\Gamma_F (\alpha_s) $ is proportional to the cusp
anomalous dimension, $\Gamma_{\mathrm{cusp}}(\alpha_s)$, which can be expanded in $\alpha_s$
\begin{equation}
\Gamma_{\mathrm{cusp}}(\alpha_s)=\sum_{n=0}^{\infty} \left( \frac{\alpha_s}{4 \pi}\right)^{1+n} \Gamma_c^n,
\end{equation}
and $\Gamma_F =(\Gamma_F^0/\Gamma_c^0)\Gamma_{\mathrm{cusp}}$. The non-cusp part, $\gamma_{F}(\alpha_s)$, has a similar expansion
\begin{equation}
\gamma_{F}(\alpha_s)=\sum_{i=0}^{\infty} \left( \frac{\alpha_s}{4 \pi}\right)^{1+i} \gamma_{F}^i.
\end{equation}
The solution to RGE is given by
\begin{equation}
F(\mu)= \exp \left( K_F (\mu, \mu_0) \right) \left( \frac{\mu_0}{m_F} \right) ^{\omega_F(\mu, \mu_0)} F(\mu_0) \, ,
\end{equation}
where the exponents $K_F$ and $\omega_F$ are given in terms of the anomalous dimension,
\begin{align}
K_F(\mu, \mu_0) &= 2 \int_{\alpha (\mu)}^{\alpha(\mu_0)} \frac{d \alpha}{\beta(\alpha)} \Gamma_F (\alpha) \int_{\alpha(\mu_0)}^{\alpha}
\frac{d \alpha'}{\beta(\alpha')} +\int_{\alpha (\mu)}^{\alpha(\mu_0)} \frac{d \alpha}{\beta(\alpha)} \gamma_F (\alpha) ,\\
\omega_F(\mu, \mu_0) &= 2 \int_{\alpha (\mu)}^{\alpha(\mu_0)} \frac{d \alpha}{\beta(\alpha)} \Gamma_F (\alpha),
\end{align}
and for up to NLL and NLL' accuracy are given by
\begin{align}
K_F(\mu, \mu_0) &=-\frac{\gamma_F^0}{2 \beta_0} \ln r -\frac{2 \pi \Gamma_F^0}{(\beta_0)^2} \Big{\lbrack} \frac{r-1+r\ln r}{\alpha_s(\mu)}
+ \left( \frac{\Gamma^1_c}{\Gamma^0_c}-\frac{\beta_1}{\beta_0} \right) \frac{1-r+\ln r}{4 \pi}+\frac{\beta_1}{8 \pi \beta_0}
\ln^2 r  \Big{\rbrack}, \\
\omega_F(\mu, \mu_0) &= - \frac{\Gamma_F^0}{j_F \beta_0} \Big{\lbrack} \ln r + \left( \frac{\Gamma^1_c}{\Gamma^0_c} -
\frac{\beta_1}{\beta_0}  \right) \frac{\alpha_s (\mu_0)}{4 \pi}(r-1)\Big{\rbrack},
\end{align}
where $r=\alpha(\mu)/\alpha(\mu_0)$ and $\beta_n$ are the coefficients of the QCD $\beta$-function,
\begin{equation}
\beta(\alpha_s) = \mu \frac{d \alpha_s}{d \mu}= -2 \alpha_s \sum_{n=0}^{\infty} \left( \frac{\alpha_s}{4 \pi} \right)^{1+n} \beta_n \, .
\end{equation}
The RGEs for functions that depend on the variable $\tau_a$ are of the form
\begin{equation}
\mu \frac{d}{d\mu} F(\tau_a, \mu) = \Big{\lbrack} \gamma_F(\mu) \otimes F(\mu) \Big{\rbrack}(\tau_a) \, ,
\end{equation}
where
\begin{align}
\begin{split}
\gamma_F(\tau_a, \mu) &= - \Big{\lbrack} Z_F^{-1} (\mu) \otimes \mu \frac{d}{d \mu} Z_F (\mu) \Big{\rbrack} (\tau_a) \\
&=\Gamma_F(\alpha_s)\left( \ln \frac{\mu^2}{m_F^2 }-\frac{2}{j_F} \left( \frac{\Theta (\tau_a)}{\tau_a} \right)_+ \right) +
\gamma_F (\alpha_s) \delta (\tau_a) \, ,
\end{split}
\end{align}
and the solution to this equation is given by
\begin{equation}
F(\tau_a, \mu) = \exp \left( K_F + \gamma_{E} \omega_F  \right) \frac{1}{\Gamma (-\omega_F)} \left(
\frac{\mu_0}{m_F} \right)^{j_F \omega_F} \Bigg{\lbrack} \left( \frac{\Theta(\tau_a)}{(\tau_a)^{1+\omega_F}}\right)_+ \otimes
F(\tau_a, \mu_0) \Bigg{\rbrack} \, .
\end{equation}
\subsection{Plus-distribution identities}
We begin  with the equation
\begin{equation}
\label{eq:plusID}
\int d\tau'' \Big{\lbrack}\frac{\Theta(\tau-\tau'')}{(\tau-\tau'')^{1+\omega_1}} \Big{\rbrack}_+ \Big{\lbrack}
\frac{\Theta(\tau''-\tau')}{(\tau''-\tau')^{1+\omega_2}} \Big{\rbrack}_+= \frac{\Gamma(-\omega_1) \Gamma(-\omega_2)}
{\Gamma(-\omega_1-\omega_2)} \Big{\lbrack}\frac{\Theta(\tau-\tau')}{(\tau-\tau')^{1+\omega_1+\omega_2}} \Big{\rbrack}_+ \, ,
\end{equation}
which can be easily proven using Laplace transforms and the defining equation of the plus distribution,
\begin{equation}
\label{eq:plusdef}
\left[f(\tau) \right]_+ \equiv \lim_{\beta \rightarrow 0} \frac{d}{d\tau}\left[ \theta(\tau-\beta) F(\tau) \right] \, ,
\end{equation}
where $F(\tau)$ is defined as
\begin{align}
F(\tau) \equiv \int^\tau_1 d \tau' f(\tau')
\,,\end{align}
which yields
\begin{equation}
{\cal{L}} \Big{\lbrace} \left( \frac{1}{\tau^{1+\omega}} \right)_+ \Big{\rbrace} = s^{\omega} \Gamma (-\omega) \, .
\end{equation}
The following equations can be derived by setting $\tau' \rightarrow 0$ in Eq.~(\ref{eq:plusID}), expanding in $\omega_2$
both sides and matching  powers:
\begin{equation}
\label{pd1}
\int d\tau' \Big{\lbrack} \frac{\Theta(\tau-\tau')}{(\tau-\tau')^{1+\omega}} \Big{\rbrack}_+ \delta(\tau') =  \Big{\lbrack}
\frac{\Theta(\tau)}{\tau^{1+\omega}} \Big{\rbrack}_+ ,
\end{equation}
\begin{equation}
\label{pd2}
\int d\tau' \Big{\lbrack} \frac{\Theta(\tau-\tau')}{(\tau-\tau')^{1+\omega}} \Big{\rbrack}_+ \Big{\lbrack} \frac{\Theta(\tau')}{\tau'}
\Big{\rbrack}_+ =  \Big{\lbrack} \frac{\Theta(\tau)}{\tau^{1+\omega}} \Big{\rbrack}_+ \left( \ln \tau - H(-1-\omega) \right), \nn
\end{equation}
\begin{equation}
\label{pd3}
\int d\tau' \Big{\lbrack} \frac{\Theta(\tau-\tau')}{(\tau-\tau')^{1+\omega}} \Big{\rbrack}_+ \Big{\lbrack} \frac{\Theta(\tau') \ln \tau' }
{\tau'} \Big{\rbrack}_+ =  \Big{\lbrack} \frac{\Theta(\tau)}{\tau^{1+\omega}} \Big{\rbrack}_+ \frac{ \left( \ln \tau - H(-1-\omega)
\right)^2+\pi^2/2 -\psi^{(1)}(-\omega)}{2} \, , \nn
\end{equation}
where we used~\cite{Ellis:2010rwa}
\begin{equation}
\Big{\lbrack} \frac{\Theta(\tau)}{\tau^{1+\omega}}  \Big{\rbrack}_+ = - \frac{1}{\omega} \delta(\tau) + \sum_{n=0}^{\infty}
(-\omega)^n  \Big{\lbrack} \frac{\Theta(\tau) \ln^n \tau}{\tau}  \Big{\rbrack}_+ .
\end{equation}

\subsection{Reorganization of logarithms of $(1-z)$}

The convolutions in the variable $z$ need to be performed numerically since they involve the evolved FFs,    which are evaluated by
solving the DGLAP equation using Mellin transformations. For this reason we expand the plus-distributions using the following
relations

\begin{equation}
\int_z^1 \frac{dx}{x}\Big{(}\frac{1}{1-x}\Big{)}_+ f \left(\frac{z}{x}\right) =\int_z^1 dx \; \frac{1}{1-x}\Big{(}\frac{1}{x}
f \left(\frac{z}{x}\right)-f(z) \Big{)} + f(z) \ln (1-z),
\end{equation}
\begin{equation}
\int_z^1 \frac{dx}{x}\Big{(}\frac{\ln (1-x)}{1-x}\Big{)}_+ f \left(\frac{z}{x}\right) =\int_z^1 dx \; \frac{\ln (1-x)}{1-x}
\Big{(}\frac{1}{x} f \left(\frac{z}{x}\right)-f(z) \Big{)} + f(z) \frac{1}{2} \ln^2 (1-z).
\end{equation}
Thus for every function $D(z)$ the convolution with $f_{\cal{J}}^{ij}(\tau, z, \mu) $ gives
\begin{align}
\begin{split}
\label{fconf}
\frac{1}{T_{ij}}\frac{2\pi}{\alpha_s(\mu)}f_{\cal{J}}^{ij}(\tau, z, \mu)\bullet D(z) &= \delta_{ij}\;f_1(\tau, z, \mu)\;D(z)-\int_{z}^{1} dx\;\; f_2(\tau, x, \mu)
\Big{(}\frac{\bar{P}_{ji}(x)}{x}\circ D\left( \frac{z}{x} \right)\Big{)}
\\
&\hspace{-0.5 in}+ \int_{z}^{1} dx \;\; \Big{\lbrack} c_{ij}(x)-\frac{1}{1-a/2} \ln \left(1+\left(\frac{1-x}{x} \right)^{1-a} \right)
\frac{\bar{P}_{ji}(x)}{x} \Big{\rbrack}\circ D\left( \frac{z}{x} \right) \,,
\end{split}
\end{align}
where
\begin{equation}
f_2(\tau, z, \mu)= 2\ln\left( \frac{\mu}{\mu_J(\tau, z)} \right) +\frac{1}{1-a/2} H(-1-\Omega) \, ,
\end{equation}
with
\begin{equation*}
\mu_J(\tau, z)=\omega \tau^{1/(2-a)}(1-z)^{(1-a)/(2-a)},
\end{equation*}
\begin{equation}
f_1(\tau, z, \mu)=\frac{1-a/2}{1-a}\Big{(}f_2(\tau, z, \mu) \Big{)}^2 + \frac{a(1-a/4)}{(1-a)(1-a/2)} \frac{\pi^2}{6} -
\frac{1}{(1-a)(1-a/2)} \psi^{(1)}(-\Omega),
\end{equation}
\begin{eqnarray*}
c_{qq}(z)&=&\frac{1-z}{z},
\\
c_{gg}(z)&=&0,
\\
c_{gq}(z)&=&2(1-z),\\
c_{qg}(z)&=&1 ,
\end{eqnarray*}
and
\begin{eqnarray*}
f(x) \circ g(x) &=& f(x)g(x)\,,\\
\lbrack f(x)(h(x))_+ \rbrack \circ g(x) &=&h(x) \lbrack f(x)g(x)-f(1)g(1) \rbrack \,.
\end{eqnarray*}

\subsection{Profile Functions}

\begin{figure}[t]
\centerline{\includegraphics[scale=0.75]{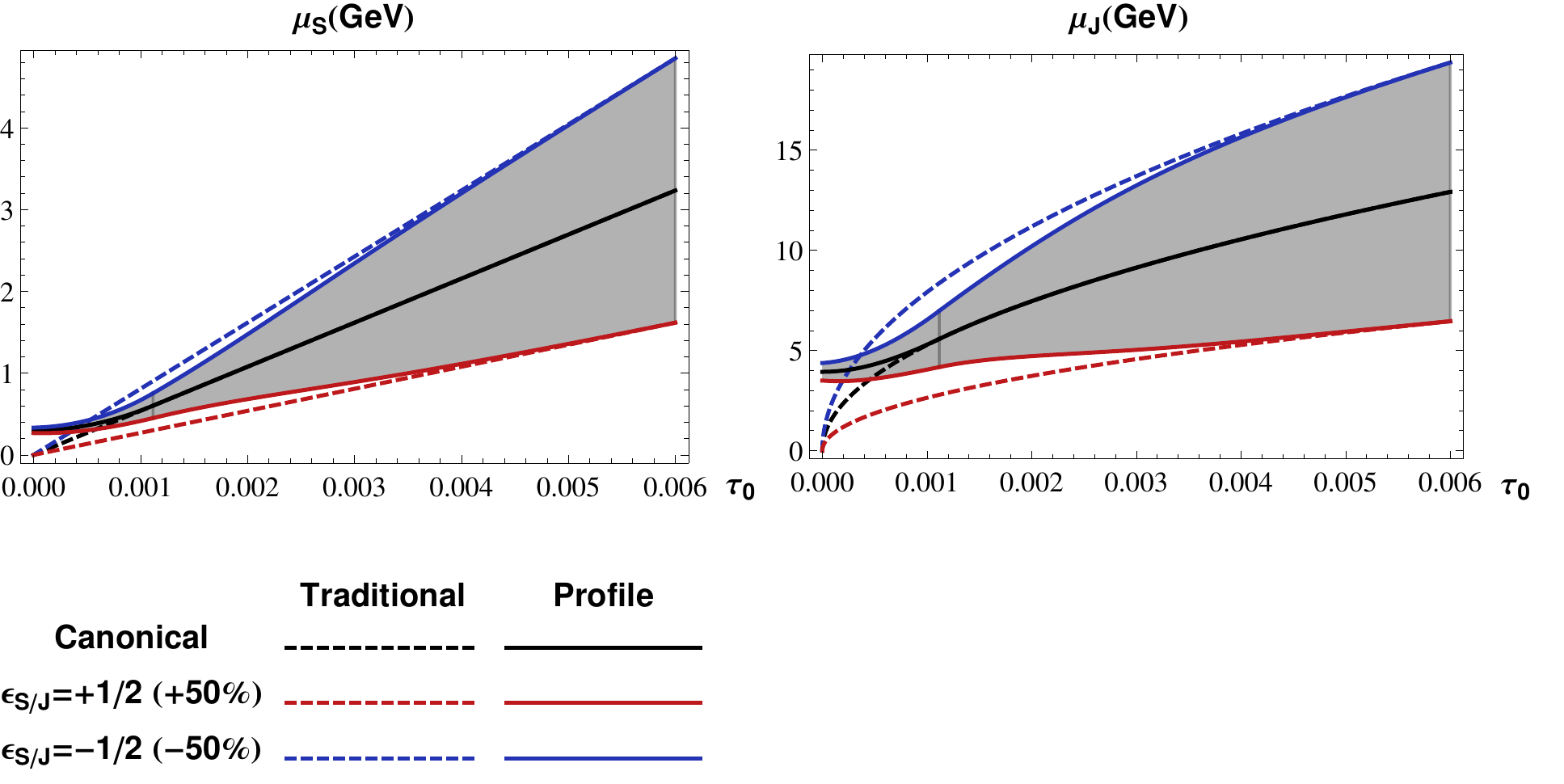}}
\caption{Profile functions for $\mu^{PF}_S(\tau_0)$ and $\mu^{PF}_J(\tau_0)$, the $\tau_0$-dependent renormalization scales that we use in the
scale variations of our measured soft function and measured jet function. Also shown are traditional scale variations done by
varying $\mu$ by $\pm 50\%$. }
\label{fig:profile_func}
\end{figure}

Here, we write down the profile functions used to perform scale variations for our measured soft and measured jet functions.
We use profile functions to introduce a $\tau_a$-dependent scale variation that freezes at the
characteristic scale for high values of $\tau_a$ where the factorization theorem breaks down and at a fixed scale for small values
of $\tau_a$ where we reach the non-perturbative regime. The profile function for the measured soft function, $\mu_S^{PF}(\tau_0)$, and
the profile function for the measured  jet function, $ \mu_J^{PF}(\tau_0)$,  are plotted in Fig.~\ref{fig:profile_func} (for the case $a=0$). The analytic formulae
for these functions are
\begin{align}
\begin{split}
  \mu_S^{PF}(\tau_a) = \left[ 1+\epsilon_S \frac{g(\tau_a)}{g(1)} \right] \times
  			\begin{cases}
               \mu_{min} + \alpha \tau_a^{\beta} \quad \quad &0 < \tau_a < \tau_{min}\\
               \omega \tau_a/ r^{(1-a)} \quad \quad &\tau_{min} \leq \tau_a\\
            \end{cases} \, , \\
            \\
  \mu_J^{PF}(\tau_a) = \left[ 1+\epsilon_J \frac{g(\tau_a)}{g(1)} \right] \times
  			\begin{cases}
               (\omega r)^{(1-a)/(2-a)}(\mu_{min}+\alpha \tau_a^{\beta})^{1/(2-a)} \quad \quad &0 < \tau_a < \tau_{min}\\
               \omega \tau_a^{1/(2-a)} \quad \quad &\tau_{min} \leq \tau_a\
            \end{cases}  \, ,
\end{split}
\end{align}
where we have defined
\begin{equation}
	g(\tau) =  \frac{1}{\exp{\Big{(} 1.26(\tau_{min}-\tau) / \tau_{min} \Big{)}} + 1}
\,,\end{equation}
and where $\alpha$ and $\beta$ are defined to be
\begin{equation}
	\beta = \frac{\tau_{min}}{\tau_{min} - \mu_{min} r^{(1-a)}/\omega} \quad \text{and}
	\quad \alpha =  \frac{\omega}{\beta \tau_{min}^{\beta-1}r^{(1-a)}} \,.
\end{equation}
These choices for $\alpha$ and $\beta$ ensure that the profile  functions and their first derivatives are continuous. We use the following values for the parameters
\begin{align}
&\tau_{min} =  2 \mu_{min} r^{1 - a}/\omega \nn\\
&\mu_{min} = 0.3 \,{\rm GeV}
\,.\end{align}

We define our scale variations via
\begin{align*}
\begin{split}
	\epsilon_{S/J} = 1/2 \quad &\to \quad +50\%  \text{\ variation}, \\
	\epsilon_{S/J} = -1/2 \quad &\to \quad -50\%  \text{\ variation}, \\
	\epsilon_{S/J} = 0 \quad &\to \quad \text{Canonical  scale}\,,
\end{split}
\end{align*}
and take the final scale variation bands as the envelope of the set of bands from the individual variations.

\section{Matching Coefficients and Consistency Checks}
\label{match}

\subsection{Evaluation of matching coefficients}
In pure dimensional regularization all diagrams contributing to the
FFs vanish, and the only diagrams that contribute to the angularity FJF for quarks  are Figs. 3a) and 3b) of Ref.~\cite{Jain:2011xz}.
For Fig. 3a) we get
\begin{multline}
\frac{C_F \alpha_s}{2\pi} \frac{(4\pi \mu^2)^\epsilon (1-\epsilon)}{\Gamma[1-\epsilon]}
\frac{1-z}{1-a/2} \,  \omega^{2 a \epsilon/(2-a)} (1-z)^{-2(1-a)\epsilon/(2-a)}\\
\times \left(1+\frac{(1-z)^{1-a}}{z^{1-a}}\right)^{2\epsilon/(2-a)}
\frac{1}{s_a^{1+2\epsilon/(2-a)}} \, ,
\end{multline}
and for Fig. 3b) we get
\begin{multline}
\frac{C_F \alpha_s}{2\pi} \frac{2 z}{1-a/2} \frac{(4\pi \mu^2)^\epsilon }{\Gamma[1-\epsilon]} \omega^{2 a \epsilon/(2-a)}
\frac{1}{(1-z)^{1+2(1-a)\epsilon/(2-a)}} \\
\times \left(1+\frac{(1-z)^{1-a}}{z^{1-a}}\right)^{2\epsilon/(2-a)} \frac{1}{s_a^{1+2\epsilon/(2-a)}} \, ,
\end{multline}
where $s_a = \omega^2\tau_a$. The first expression is singular as $\tau_a \to 0$ the second is singular as $z\to 1$ and $\tau_a \to 0$, but the singularities
are regulated by dimensional regularization. Employing  the distributional
identity
\bea
\frac{1}{(1-z)^{1+\epsilon} }= - \frac{1}{\epsilon} \delta(1-z) +\left( \frac{1}{1-z}\right)_+ - \epsilon \left(
\frac{\ln(1-z)}{1-z}\right)_+ + ... \, ,
\eea
and similarly for $\tau_a$ we find for the divergent terms
\begin{multline}
 \frac{C_F \alpha_s}{2\pi} \Bigg{(} \delta(s_a) \delta(1-z) \left[  \frac{2-a}{1-a}\frac{1}{\epsilon^2} + \frac{2-a}{1-a}
 \frac{1}{\epsilon}\ln\left(\frac{\mu^2}{\omega^2}\right) +\frac{3}{2\epsilon} \right] \\
 -\frac{1}{1-a}\frac{2}{\epsilon} \delta(1-z) \frac{1}{\omega^2} \left[\frac{1}{\tau_a}\right]_+ - \delta(s_a)
 \frac{1}{\epsilon} P_{qq}(z) \Bigg{)} \, ,
\end{multline}
where $P_{qq}$ is defined in Eq.~(\ref{eq:split}). The first four terms in this expression are the expected UV poles for the angularity jet function (multiplied by $\delta(1-z)$),
see Eq.~(3.37) of Ref.~\cite{Hornig:2009vb}. In order to simplify this  expression we have redefined  $4\pi e^{-\gamma_E} \mu^2 \to \mu^2$, i.e.,
we are working in the $\overline{MS}$ scheme.
The last term is the expected UV pole in the perturbative  evaluation of the QCD fragmentation function. Since
${\cal G}_i(\tau_a,z,\mu)$ is expected to evolve like the angularity jet function,
this is the correct structure of UV divergences implied by Eq.~(\ref{eq:GintermsofD}).
The finite pieces are given by
\begin{align}
\label{eq:Jqq}
\frac{1}{\omega^2}\frac{{\cal J}_{qq}(\tau_a,z,\mu)}{2 (2\pi)^3} &= \frac{ C_F \alpha_s}{2\pi} \frac{1}{\omega^2}
\left\{ \delta(\tau_a)\delta(1-z)\frac{2-a}{1-a} \left(  - \frac{\pi^2}{12} +\frac{1}{2}\ln^2\left(
\frac{\mu^2}{\omega^2} \right) \right)   \right. \nn\\
&+\delta(\tau_a)\left( 1-z -\left[\ln\left( \frac{\mu^2}{\omega^2} \right) + \frac{1}{1-a/2}\ln\left(1+\frac{(1-z)^{1-a}}{z^{1-a}}\right)
\right]  \frac{1+z^2}{(1-z)_+}  \right. \nn\\
&+ \left. \frac{1-a}{1-a/2}  (1+z^2)\left(\frac{\ln(1-z)}{1-z}\right)_+  \right)  \nn\\
&+   \left[\frac{1}{\tau_a}\right]_+\left(\frac{1}{1-a/2} \frac{1+z^2}{(1-z)_+} -\delta(1-z)\frac{2}{1-a}\ln\left(
\frac{\mu^2}{\omega^2}\right) \right) \nn\\
&\left.+ \frac{2\delta(1-z)}{(1-a)(1-a/2)} \left[\frac{\ln \tau_a}{\tau_a}\right]_+\right\} \,.
\end{align}
In the limit $a\to0$ this becomes
\begin{align}
\begin{split}
\frac{1}{\omega^2}\frac{{\cal J}_{qq}(\tau_0,z,\mu)}{2 (2\pi)^3}  &=\frac{ C_F \alpha_s}{2\pi}  \left\{ \delta(s)\delta(1-z)
 \left( - \frac{\pi^2}{6} + \ln^2\left(\frac{\mu^2}{\omega^2} \right) \right)   \right. \\
&+\delta(s)\left( 1-z - \ln\left( \frac{\mu^2}{\omega^2} \right) \frac{1+z^2}{(1-z)_+} + \ln z\, P_{qq}(z) + (1+z^2)
\left(\frac{\ln(1-z)}{1-z}\right)_+   \right) \\
&\left. + \frac{1}{\omega^2} \left[\frac{1}{\tau_0}\right]_+\left( \frac{1+z^2}{(1-z)_+} -2 \delta(1-z)\ln\left(
 \frac{\mu^2}{\omega^2}\right)  \right)+2\delta(1-z) \frac{1}{\omega^2}\left[\frac{\ln \tau_0}{\tau_0}\right]_+ \right\} \, ,
\end{split}
\end{align}
where we have used $\delta(\tau_0)/\omega^2 = \delta(s)$.
Using the following distributional identities
\begin{align}
\begin{split}
\frac{1}{\omega^2}\left[\frac{1}{\tau_0}\right]_+ &= \frac{1}{\omega^2}\left[\frac{\omega^2}{s}\right]_+ =
\frac{1}{\mu^2}\left[ \frac{\mu^2}{s}\right]_+  + \ln\left(\frac{\mu^2}{\omega^2}\right) \delta(s)\,, \\
\frac{1}{\omega^2}\left[\frac{\ln \tau_0}{\tau_0}\right]_+ &= \frac{1}{\omega^2}\left[\frac{\ln(s/\omega^2)}{s/\omega^2}\right]_+ =
\frac{1}{\mu^2}\left[\frac{\ln(s/\mu^2)}{s/\mu^2}\right]_+ +  \frac{\ln(\mu^2/\omega^2)}{\mu^2}\left[ \frac{\mu^2}{s}\right]_+ +
\frac{1}{2} \ln\left(\frac{\mu^2}{\omega^2}\right) \delta(s) \, ,
\end{split}
\end{align}
which are readily verified by integrating both sides over $s$, one finds that in the $a \to 0$ limit the finite piece is given by
\begin{align}
\frac{{\cal J}_{qq}(s,z,\mu)}{2(2 \pi)^3)} &= \frac{ C_F \alpha_s}{2\pi}   \left\{ \delta(s)\left( 1-z  + \ln z\, P_{qq}(z) +
(1+z^2)\left(\frac{\ln(1-z)}{1-z}\right)_+   - \frac{\pi^2}{6} \delta(1-z)\right) \right.  \nn\\
&\left. + \frac{1}{\mu^2} \left[\frac{\mu^2}{s}\right]_+ \frac{1+z^2}{(1-z)_+}  +2\delta(1-z) \frac{1}{\mu^2}
\left[\frac{\ln(s/\mu^2)}{s/\mu^2}\right]_+ \right\} \, ,
\end{align}
which agrees with  the matching coefficient  found in Eq.~(2.32) of Ref.~\cite{Jain:2011xz}.

Next we calculate ${\cal J}_{qg}(\tau_a,z,\mu)$. Naively this is related to ${\cal J}_{qq}(\tau_a,z,\mu)$ by the replacement $z\to1-z$.
However, because in the convolution integral of Eq.~(\ref{eq:GintermsofD}) the argument of ${\cal J}_{ij}(\tau_a,z/z',\mu)$ is never zero, there is no need to
regulate poles of $z$. Therefore, a divergent factor of $(1-z)^{-1-\epsilon}$ in ${\cal J}_{qq}(\tau_a,z, \mu)$ becomes in
${\cal J}_{qg}(\tau_a,z,\mu)$
\begin{equation}
\frac{1}{z^{1+\epsilon}}= \frac{1}{z} - \epsilon \frac{\ln z}{z} + O(\epsilon^2)\,.
\end{equation}
Thus,  ${\cal J}_{qg}(\tau_a,z,\mu)$ is obtained by making the substitution $z \to 1-z$ and then dropping all $\delta(z)$
and plus prescriptions.
This is true for the ${\cal J}_{qg}(s,z,\mu)$ calculated in Ref.~\cite{Jain:2011xz} and remains true for ${\cal J}_{qg}(\tau_a,z,\mu)$. We
thus find for the divergent
terms
\begin{equation}
\frac{1}{\omega^2}\frac{{\cal J}^{div}_{qg}(\tau_a,z, \mu) }{2 (2\pi)^3} =  - \frac{1}{\omega^2}\frac{C_F \alpha_s}{2\pi} \frac{1}{\epsilon}
\delta(\tau_a) P_{gq}(z) \, ,
\end{equation}
where $P_{gq}$ is given in Eq.~(\ref{eq:split}). For the finite pieces we get
\begin{align}
\begin{split}
\frac{1}{\omega^2}\frac{{\cal J}_{qg}(\tau_a,z, \mu) }{2 (2\pi)^3} &=  \frac{C_F \alpha_s}{2\pi} \frac{1}{\omega^2}
\Bigg{\{} \delta(\tau_a) \Big{(} z +\Big{\lbrack} \frac{1}{1-a/2}\ln\left(\frac{z^{1-a}(1-z)^{1-a}}{z^{1-a}+(1-z)^{1-a}}\right) \\
&- \ln\left(\frac{\mu^2}{\omega^2} \right)  \Big{\rbrack} P_{gq}(z) \Big{)}
+ \frac{1}{1-a/2}\left[\frac{1}{\tau_a}\right]_+ P_{gq}(z)  \Bigg{\}} \,.
\end{split}
\end{align}
Again, these reproduce the matching coefficients of Ref.~\cite{Jain:2011xz} in the $a\to 0$ limit.

For the divergent contributions to ${\cal J}_{gg}(\tau_a,z,\mu)$ we get (from the diagrams in Fig.~4 of Ref.~\cite{Jain:2011xz})
\begin{align}
\frac{1}{\omega^2}\frac{\mathcal{J}^{div}_{gg}(\tau_a,z,\mu)}{2 (2\pi)^3} &= \frac{C_A \alpha_s}{2\pi} \frac{1}{\omega^2} \Bigg{(}
\delta(\tau_a) \delta(1-z) \left[  \frac{2-a}{1-a}\frac{1}{\epsilon^2} + \frac{2-a}{1-a}\frac{1}{\epsilon}\ln \left( \frac{\mu^2}{\omega^2}
\right) +\frac{\beta_0}{2 C_A}\frac{1}{\epsilon} \right]  \nn\\
&-\frac{1}{1-a}\frac{2}{\epsilon} \delta(1-z)\left[\frac{1}{\tau_a}\right]_+  \Bigg{)} -  \frac{\alpha_s}{2\pi} \frac{1}{\omega^2}
\delta(\tau_a) \frac{1}{\epsilon} \tilde{P}_{gg}(z) \, ,
\end{align}
where the $\tilde{P}_{gg}(z)$ is the ${\bf full}$ QCD splitting function that includes the term proportional to $\beta_0 \delta(1-z)$.
For the finite parts of ${\cal J}_{gg}(\tau_a,z,\mu)$ we find
\bea
\frac{1}{\omega^2}\frac{{\cal J}_{gg}(\tau_a,z,\mu)}{2 (2\pi)^3} &=& \frac{ C_A \alpha_s}{2\pi} \frac{1}{\omega^2}
\left\{ \delta(\tau_a)\delta(1-z)\frac{2-a}{1-a} \left(  - \frac{\pi^2}{12} +\frac{1}{2}\ln^2\left(\frac{\mu^2}{\omega^2}
\right) \right)   \right. \nn\\
&&+\delta(\tau_a)\left(- P_{gg}(z)\left[\ln\left( \frac{\mu^2}{\omega^2} \right) + \frac{1}{1-a/2}
\ln\left(1+\frac{(1-z)^{1-a}}{z^{1-a}}\right)\right]   \right. \nn \\
&&+ \left. \frac{1-a}{1-a/2}  \frac{2(1-z+z^2)^2}{z}\left(\frac{\ln(1-z)}{1-z}\right)_+  \right)  \nn \\
&&+   \left[\frac{1}{\tau_a}\right]_+\left(\frac{1}{1-a/2} P_{gg}(z) -\delta(1-z)\frac{2}{1-a}
\ln\left(  \frac{\mu^2}{\omega^2}\right) \right) \nn\\
&&\left.+ \frac{2\delta(1-z)}{(1-a)(1-a/2)} \left[\frac{\ln \tau_a}{\tau_a}\right]_+\right\} \, ,
\eea
where
$P_{gg}$ is given in Eq.~(\ref{eq:split}).
In the limit $a\to 0$, this expression reduces to  ${\cal J}_{gg}(s,z,\mu)/(16 \pi^3)$ found in Eq.~(2.33) of Ref.~\cite{Jain:2011xz}.

For  the divergent contributions to ${\cal J}_{gq}(\tau_a,z,\mu)$ we find
\bea
\frac{1}{\omega^2}\frac{{\cal J}^{div}_{gq}(\tau_a,z,\mu)}{2 (2\pi)^3} &=& -\frac{1}{\omega^2} \frac{\alpha_s T_R}{2\pi} \frac{1}{\epsilon}
 \delta(\tau_a) P_{qg}(z) \, .
 \eea
For the finite parts we get
\bea
\frac{1}{\omega^2}\frac{{\cal J}_{gq}(\tau_a,z,\mu)}{2 (2\pi)^3} &=& \frac{\alpha_s T_R}{2\pi} \frac{1}{\omega^2}
\left\{ \frac{1}{1-a/2} \left[ \frac{1}{\tau_a}\right]_+ P_{qg}(z)  +  \delta(\tau_a) 2z(1-z)
\right.  \\
&&\left. + \,\delta(\tau_a) P_{qg}(z) \left[  \frac{1}{1-a/2}\ln\left(\frac{z^{1-a}(1-z)^{1-a}}{z^{1-a}+(1-z)^{1-a}}\right)  -
\ln\left( \frac{\mu^2}{\omega^2} \right) \right]
\right\} \nn  \, ,
\eea
where $P_{qg}$ is again given in Eq.~(\ref{eq:split}). In the limit $a\to 0$, this expression reduces to ${\cal J}_{gq}(s,z,\mu)/(16 \pi^3)$ in Eq.~(2.33) of Ref.~\cite{Jain:2011xz}.
\subsection{Sum Rules}
The sum rules,
\begin{align}
J_{i}(\tau_a) &= \frac{1}{2(2\pi)^3} \sum_j \int_0^1 dz \,z \, \mathcal{J}_{ij}(\tau_a, z) \, ,
\end{align}
can be checked for $i=q$ by performing the integral
\begin{align}
J_{q}(\tau_a) &= \frac{1}{2(2\pi)^3} \sum_j \int_0^1 dz \,z \, {\cal{J}}_{qj}(\tau_a, z) \\
&=\frac{1}{2(2\pi)^3} \int_0^1 dz \,z \, \left(\mathcal{J}_{qq}(\tau_a, z)+ \mathcal{J} _{qg}(\tau_a, z)\right) \\
&=\frac{1}{2(2\pi)^3} \int_0^1 dz \,z \, \left(\mathcal{J}_{qq}(\tau_a, z)+ \mathcal{J} _{qq}(\tau_a, 1-z)\right) \\
&=\frac{1}{2(2\pi)^3} \int_0^1 dz \, \mathcal{J}_{qq}(\tau_a, z),
\end{align}
where in the last line we changed variables to $z \to 1-z$ in the 2nd term. Inserting the expression in Eq.~(\ref{eq:Jqq}) into this integral yields
the $J_{q}(\tau_a)$ found in Eq.~(3.35) of Ref.~\cite{Hornig:2009vb}.

In the case of the $i=g$ we have
\begin{align}\label{eq:gsumrule}
J_{g}(\tau_a) &= \frac{1}{2(2\pi)^3}  \int_0^1 dz \,z \,  \left(\mathcal{J}_{gg}(\tau_a, z)+ \mathcal{J} _{gq}(\tau_a, z)\right) \, \nn \\
 &= \frac{1}{2(2\pi)^3}   \int_0^1 dz  \,  \frac{  \mathcal{J}_{gg}(\tau_a, z)+ \mathcal{J} _{gq}(\tau_a, z)}{2} \, ,
\end{align}
because both $\mathcal{J}_{gg}(\tau_a, z)$ and $\mathcal{J} _{gq}(\tau_a, z)$ are symmetric under $z \to 1-z$.
The sum rule is easiest to verify by writing the $d$-dimensional expressions for $\mathcal{J}_{gg}(\tau_a, z)$ and $\mathcal{J} _{gq}(\tau_a, z)$
before expanding in $\epsilon = (4-d)/2$. We find
\bea
\frac{1}{\omega^2} \frac{\mathcal{J}_{gg}(\tau_a, z,\mu)}{2(2\pi)^3} &=&\frac{1}{\omega^2} \left(\frac{4\pi \mu^2}{\omega^2}\right)^\epsilon \frac{C_A \alpha_s}{2\pi}\frac{1}{\Gamma[1-\epsilon]}
\frac{1}{1-a/2} (z^{a-1} + (1-z)^{a-1})^{\frac{2\epsilon}{2-a}} \left(\frac{1}{\tau_a}\right)^{1+\frac{2\epsilon}{1-a}} \nn \\
&\times&\left( \frac{2 z}{1-z} +\frac{2 (1-z)}{z} +2 z(1-z)\right) \\
\frac{1}{\omega^2} \frac{\mathcal{J}_{gq}(\tau_a, z,\mu)}{2(2\pi)^3} &=&\frac{1}{\omega^2} \left(\frac{4\pi \mu^2}{\omega^2}\right)^\epsilon \frac{T_R \alpha_s}{2\pi}\frac{1}{\Gamma[1-\epsilon]}
\frac{1}{1-a/2} (z^{a-1} + (1-z)^{a-1})^{\frac{2\epsilon}{2-a}} \left(\frac{1}{\tau_a}\right)^{1+\frac{2\epsilon}{1-a}} \nn \\
&\times&\left(1 - \frac{2}{1-\epsilon}  z(1-z)\right) \,.
\eea
Inserting these two expressions into Eq.~(\ref{eq:gsumrule}) one obtains exactly the integral expression for the $d$-dimensional $J_g(\tau_a)$ found in Eq.~(4.22) of
Ref.~\cite{Ellis:2010rwa}.

\bibliography{paper}

\providecommand{\href}[2]{#2}\begingroup\raggedright\begin{thebibliography}{10}

\bibitem{Almeida:2014uva}
L.~G. Almeida, S.~D. Ellis, C.~Lee, G.~Sterman, I.~Sung and J.~R. Walsh,
  \emph{{Comparing and counting logs in direct and effective methods of QCD
  resummation}}, \href{http://dx.doi.org/10.1007/JHEP04(2014)174}{\emph{JHEP}
  {\bf 04} (2014) 174}, [\href{http://arxiv.org/abs/1401.4460}{{\tt
  1401.4460}}].

\bibitem{Procura:2009vm}
M.~Procura and I.~W. Stewart, \emph{{Quark Fragmentation within an Identified
  Jet}}, \href{http://dx.doi.org/10.1103/PhysRevD.81.074009,
  10.1103/PhysRevD.83.039902}{\emph{Phys.Rev.} {\bf D81} (2010) 074009},
  [\href{http://arxiv.org/abs/0911.4980}{{\tt 0911.4980}}].

\bibitem{Liu:2010ng}
X.~Liu, \emph{{SCET approach to top quark decay}},
  \href{http://dx.doi.org/10.1016/j.physletb.2011.03.055}{\emph{Phys.Lett.}
  {\bf B699} (2011) 87--92}, [\href{http://arxiv.org/abs/1011.3872}{{\tt
  1011.3872}}].

\bibitem{Jain:2011xz}
A.~Jain, M.~Procura and W.~J. Waalewijn, \emph{{Parton Fragmentation within an
  Identified Jet at NNLL}},
  \href{http://dx.doi.org/10.1007/JHEP05(2011)035}{\emph{JHEP} {\bf 1105}
  (2011) 035}, [\href{http://arxiv.org/abs/1101.4953}{{\tt 1101.4953}}].

\bibitem{Jain:2011iu}
A.~Jain, M.~Procura and W.~J. Waalewijn, \emph{{Fully-Unintegrated Parton
  Distribution and Fragmentation Functions at Perturbative $k_T$}},
  \href{http://dx.doi.org/10.1007/JHEP04(2012)132}{\emph{JHEP} {\bf 1204}
  (2012) 132}, [\href{http://arxiv.org/abs/1110.0839}{{\tt 1110.0839}}].

\bibitem{Procura:2011aq}
M.~Procura and W.~J. Waalewijn, \emph{{Fragmentation in Jets: Cone and
  Threshold Effects}},
  \href{http://dx.doi.org/10.1103/PhysRevD.85.114041}{\emph{Phys.Rev.} {\bf
  D85} (2012) 114041}, [\href{http://arxiv.org/abs/1111.6605}{{\tt
  1111.6605}}].

\bibitem{Jain:2012uq}
A.~Jain, M.~Procura, B.~Shotwell and W.~J. Waalewijn, \emph{{Fragmentation with
  a Cut on Thrust: Predictions for B-factories}},
  \href{http://dx.doi.org/10.1103/PhysRevD.87.074013}{\emph{Phys. Rev.} {\bf
  D87} (2013) 074013}, [\href{http://arxiv.org/abs/1207.4788}{{\tt
  1207.4788}}].

\bibitem{Bauer:2013bza}
C.~W. Bauer and E.~Mereghetti, \emph{{Heavy Quark Fragmenting Jet Functions}},
  \href{http://dx.doi.org/10.1007/JHEP04(2014)051}{\emph{JHEP} {\bf 04} (2014)
  051}, [\href{http://arxiv.org/abs/1312.5605}{{\tt 1312.5605}}].

\bibitem{Baumgart:2014upa}
M.~Baumgart, A.~K. Leibovich, T.~Mehen and I.~Z. Rothstein, \emph{{Probing
  Quarkonium Production Mechanisms with Jet Substructure}},
  \href{http://dx.doi.org/10.1007/JHEP11(2014)003}{\emph{JHEP} {\bf 11} (2014)
  003}, [\href{http://arxiv.org/abs/1406.2295}{{\tt 1406.2295}}].

\bibitem{Kaufmann:2015hma}
T.~Kaufmann, A.~Mukherjee and W.~Vogelsang, \emph{{Hadron Fragmentation Inside
  Jets in Hadronic Collisions}},
  \href{http://dx.doi.org/10.1103/PhysRevD.92.054015}{\emph{Phys. Rev.} {\bf
  D92} (2015) 054015}, [\href{http://arxiv.org/abs/1506.01415}{{\tt
  1506.01415}}].

\bibitem{Chien:2015ctp}
Y.-T. Chien, Z.-B. Kang, F.~Ringer, I.~Vitev and H.~Xing, \emph{{Jet
  fragmentation functions in proton-proton collisions using soft-collinear
  effective theory}},  \href{http://arxiv.org/abs/1512.06851}{{\tt
  1512.06851}}.

\bibitem{Bodwin:1994jh}
G.~T. Bodwin, E.~Braaten and G.~P. Lepage, \emph{{Rigorous QCD analysis of
  inclusive annihilation and production of heavy quarkonium}},
  \href{http://dx.doi.org/10.1103/PhysRevD.55.5853,
  10.1103/PhysRevD.51.1125}{\emph{Phys. Rev.} {\bf D51} (1995) 1125--1171},
  [\href{http://arxiv.org/abs/hep-ph/9407339}{{\tt hep-ph/9407339}}].

\bibitem{Berger:2003iw}
C.~F. Berger, T.~Kucs and G.~Sterman, \emph{Event shape / energy flow
  correlations}, {\emph{Phys. Rev.} {\bf D68} (2003) 014012},
  [\href{http://arxiv.org/abs/hep-ph/0303051}{{\tt hep-ph/0303051}}].

\bibitem{Alwall:2014hca}
J.~Alwall, R.~Frederix, S.~Frixione, V.~Hirschi, F.~Maltoni, O.~Mattelaer
  et~al., \emph{{The automated computation of tree-level and next-to-leading
  order differential cross sections, and their matching to parton shower
  simulations}}, \href{http://dx.doi.org/10.1007/JHEP07(2014)079}{\emph{JHEP}
  {\bf 07} (2014) 079}, [\href{http://arxiv.org/abs/1405.0301}{{\tt
  1405.0301}}].

\bibitem{Sjostrand:2006za}
T.~Sjostrand, S.~Mrenna and P.~Z. Skands, \emph{{PYTHIA 6.4 Physics and
  Manual}}, \href{http://dx.doi.org/10.1088/1126-6708/2006/05/026}{\emph{JHEP}
  {\bf 05} (2006) 026}, [\href{http://arxiv.org/abs/hep-ph/0603175}{{\tt
  hep-ph/0603175}}].

\bibitem{Sjostrand:2014zea}
\emph{{An Introduction to PYTHIA 8.2}},
  \href{http://dx.doi.org/10.1016/j.cpc.2015.01.024}{\emph{Comput. Phys.
  Commun.} {\bf 191} (2015) 159--177},
  [\href{http://arxiv.org/abs/1410.3012}{{\tt 1410.3012}}].

\bibitem{Bahr:2008pv}
M.~Bahr et~al., \emph{{Herwig++ Physics and Manual}},
  \href{http://dx.doi.org/10.1140/epjc/s10052-008-0798-9}{\emph{Eur. Phys. J.}
  {\bf C58} (2008) 639--707}, [\href{http://arxiv.org/abs/0803.0883}{{\tt
  0803.0883}}].

\bibitem{Ellis:2010rwa}
S.~D. Ellis, C.~K. Vermilion, J.~R. Walsh, A.~Hornig and C.~Lee, \emph{{Jet
  Shapes and Jet Algorithms in SCET}},
  \href{http://dx.doi.org/10.1007/JHEP11(2010)101}{\emph{JHEP} {\bf 11} (2010)
  101}, [\href{http://arxiv.org/abs/1001.0014}{{\tt 1001.0014}}].

\bibitem{Chien:2015cka}
Y.-T. Chien, A.~Hornig and C.~Lee, \emph{{A Soft-Collinear Mode for Jet Cross
  Sections in Soft Collinear Effective Theory}},
  \href{http://arxiv.org/abs/1509.04287}{{\tt 1509.04287}}.

\bibitem{Kartvelishvili:1978jh}
V.~G. Kartvelishvili and A.~K. Likhoded, \emph{{Heavy Quark Fragmentation Into
  Mesons and Baryons}}, {\emph{Sov. J. Nucl. Phys.} {\bf 29} (1979) 390}.

\bibitem{Kniehl:2008zza}
B.~A. Kniehl, G.~Kramer, I.~Schienbein and H.~Spiesberger, \emph{{Finite-mass
  effects on inclusive $B$ meson hadroproduction}},
  \href{http://dx.doi.org/10.1103/PhysRevD.77.014011}{\emph{Phys. Rev.} {\bf
  D77} (2008) 014011}, [\href{http://arxiv.org/abs/0705.4392}{{\tt
  0705.4392}}].

\bibitem{Cacciari:2011ma}
M.~Cacciari, G.~P. Salam and G.~Soyez, \emph{{FastJet User Manual}},
  \href{http://dx.doi.org/10.1140/epjc/s10052-012-1896-2}{\emph{Eur. Phys. J.}
  {\bf C72} (2012) 1896}, [\href{http://arxiv.org/abs/1111.6097}{{\tt
  1111.6097}}].

\bibitem{Ligeti:2008ac}
Z.~Ligeti, I.~W. Stewart and F.~J. Tackmann, \emph{Treating the {$b$} quark
  distribution function with reliable uncertainties},
  \href{http://dx.doi.org/10.1103/PhysRevD.78.114014}{\emph{Phys. Rev.} {\bf
  D78} (2008) 114014}, [\href{http://arxiv.org/abs/0807.1926}{{\tt
  0807.1926}}].

\bibitem{Abbate:2010xh}
R.~Abbate, M.~Fickinger, A.~H. Hoang, V.~Mateu and I.~W. Stewart, \emph{{Thrust
  at $N^3LL$ with Power Corrections and a Precision Global Fit for
  alphas(mZ)}},
  \href{http://dx.doi.org/10.1103/PhysRevD.83.074021}{\emph{Phys.Rev.} {\bf
  D83} (2011) 074021}, [\href{http://arxiv.org/abs/1006.3080}{{\tt
  1006.3080}}].

\bibitem{Hornig:2016ahz}
A.~Hornig, Y.~Makris and T.~Mehen, \emph{{Jet Shapes in Dijet Events at the LHC
  in SCET}},  \href{http://arxiv.org/abs/1601.01319}{{\tt 1601.01319}}.

\bibitem{Braaten:1993mp}
E.~Braaten, K.-m. Cheung and T.~C. Yuan, \emph{{Z0 decay into charmonium via
  charm quark fragmentation}},
  \href{http://dx.doi.org/10.1103/PhysRevD.48.4230}{\emph{Phys. Rev.} {\bf D48}
  (1993) 4230--4235}, [\href{http://arxiv.org/abs/hep-ph/9302307}{{\tt
  hep-ph/9302307}}].

\bibitem{Braaten:1993rw}
E.~Braaten and T.~C. Yuan, \emph{{Gluon fragmentation into heavy quarkonium}},
  \href{http://dx.doi.org/10.1103/PhysRevLett.71.1673}{\emph{Phys. Rev. Lett.}
  {\bf 71} (1993) 1673--1676}, [\href{http://arxiv.org/abs/hep-ph/9303205}{{\tt
  hep-ph/9303205}}].

\bibitem{Braaten:1994vv}
E.~Braaten and S.~Fleming, \emph{{Color octet fragmentation and the psi-prime
  surplus at the Tevatron}},
  \href{http://dx.doi.org/10.1103/PhysRevLett.74.3327}{\emph{Phys. Rev. Lett.}
  {\bf 74} (1995) 3327--3330}, [\href{http://arxiv.org/abs/hep-ph/9411365}{{\tt
  hep-ph/9411365}}].

\bibitem{Braaten:1996jt}
E.~Braaten and Y.-Q. Chen, \emph{{Helicity decomposition for inclusive J / psi
  production}}, \href{http://dx.doi.org/10.1103/PhysRevD.54.3216}{\emph{Phys.
  Rev.} {\bf D54} (1996) 3216--3227},
  [\href{http://arxiv.org/abs/hep-ph/9604237}{{\tt hep-ph/9604237}}].

\bibitem{Butenschoen:2011yh}
M.~Butenschoen and B.~A. Kniehl, \emph{{World data of J/psi production
  consolidate NRQCD factorization at NLO}},
  \href{http://dx.doi.org/10.1103/PhysRevD.84.051501}{\emph{Phys. Rev.} {\bf
  D84} (2011) 051501}, [\href{http://arxiv.org/abs/1105.0820}{{\tt
  1105.0820}}].

\bibitem{Butenschoen:2012qr}
M.~Butenschoen and B.~A. Kniehl, \emph{{Next-to-leading-order tests of NRQCD
  factorization with $J/\psi$ yield and polarization}},
  \href{http://dx.doi.org/10.1142/S0217732313500272}{\emph{Mod. Phys. Lett.}
  {\bf A28} (2013) 1350027}, [\href{http://arxiv.org/abs/1212.2037}{{\tt
  1212.2037}}].

\bibitem{Buckley:2010ar}
A.~Buckley, J.~Butterworth, L.~Lonnblad, D.~Grellscheid, H.~Hoeth, J.~Monk
  et~al., \emph{{Rivet user manual}},
  \href{http://dx.doi.org/10.1016/j.cpc.2013.05.021}{\emph{Comput. Phys.
  Commun.} {\bf 184} (2013) 2803--2819},
  [\href{http://arxiv.org/abs/1003.0694}{{\tt 1003.0694}}].

\bibitem{Lee:2006nr}
C.~Lee and G.~Sterman, \emph{Momentum flow correlations from event shapes:
  Factorized soft gluons and {Soft-Collinear Effective Theory}}, {\emph{Phys.
  Rev.} {\bf D75} (2007) 014022},
  [\href{http://arxiv.org/abs/hep-ph/0611061}{{\tt hep-ph/0611061}}].

\bibitem{Hornig:2009vb}
A.~Hornig, C.~Lee and G.~Ovanesyan, \emph{Effective predictions of event
  shapes: Factorized, resummed, and gapped angularity distributions},
  \href{http://dx.doi.org/10.1088/1126-6708/2009/05/122}{\emph{JHEP} {\bf 05}
  (2009) 122}, [\href{http://arxiv.org/abs/0901.3780}{{\tt 0901.3780}}].

\end{thebibliography}\endgroup

\end{document}